\documentclass{article}
\usepackage[a4paper, total={6in, 8in}]{geometry}
\usepackage[utf8]{inputenc}
\usepackage{amsmath, amsfonts,amsthm}
\usepackage{float}
\usepackage{cite}
\usepackage{graphicx}
\usepackage{textcomp}
\usepackage{pgfplots}
\usepackage[export]{adjustbox}
\usetikzlibrary{spy}
\usepackage{graphicx}
\usepackage{tabularx}
\usepackage{amssymb,amsmath,amsfonts, amsthm}
\usepackage{tikz}
\usepackage{multirow}
\usepackage{booktabs}
\usepackage{blindtext}
\usepackage{multicol}
\usepackage{tikzscale}
\usepackage{numprint}
\npdecimalsign{.}
\nprounddigits{3} 
\usepackage{caption}
\usepackage{subcaption}
\usepackage{parskip}
\usepackage{enumitem}
\usepackage{float}
\usepackage{pict2e}
\usepackage{siunitx}
\usepackage{xcolor}
\usepackage[english]{babel}
\usepackage{ulem}
\usepackage{authblk}
\newcommand{\C}{\mathbb C}

\newcommand{\R}{\mathbb R}

\newcommand{\Rd}{\mathbf{R}}

\newcommand{\sd}{\mathbf{s}}

\newcommand{\ZZ}{\mathbf{z}}

\newcommand{\Fd}{\mathbf F} 
                       
\newcommand{\Id}{\mathbf I}                        
\newcommand{\Cd}{\mathbf C}                        
\newcommand{\Ed}{\mathbf E}

\newcommand{\Eu}{\mathbf{E}_I}

\newcommand{\herm}{{\scriptstyle \boldsymbol{\mathsf{H}}}}
\newcommand{\trans}{{\scriptstyle \boldsymbol{\mathsf{T}}}}

\newcommand\bigcircle{\raisebox{-0.5mm}{\scalebox{1.7}{$\bigcirc$}}}

\usepackage{colortbl}

\usepackage{mathtools}

\usepackage{mathtools}
\usepackage[percent]{overpic}
\usepackage{algorithm,algcompatible,amsmath}
\algnewcommand\INPUT{\item[\textbf{Input:}]}%
\algnewcommand\PARAMETER{\item[\textbf{Parameters:}]}%
\algnewcommand\OUTPUT{\item[\textbf{Output:}]}%
\usepackage[
 colorlinks=false, backref=page
]{hyperref}
\renewcommand*{\backref}[1]{}
\renewcommand*{\backrefalt}[4]{%
    \ifcase #1 (Not cited).%
    \or        (Cited on page~#2).%
    \else      (Cited on pages~#2).%
    \fi}

\colorlet{lred}{red!80}
\colorlet{cmix}{blue!80!red} 
\colorlet{lgreen}{green!80}
\colorlet{lblue}{blue!80}

\newcommand{\stkout}[1]{\ifmmode\text{\sout{\ensuremath{#1}}}\else\sout{#1}\fi}
\usepackage{algorithmicx}
\usepackage{comment}

\usepackage{array}

\newcolumntype{L}[1]{>{\raggedright\let\newline\\\arraybackslash\hspace{0pt}}m{#1}}
\newcolumntype{C}[1]{>{\centering\let\newline\\\arraybackslash\hspace{0pt}}m{#1}}
\newcolumntype{R}[1]{>{\raggedleft\let\newline\\\arraybackslash\hspace{0pt}}m{#1}}

\title{\vspace{-1em}MR imaging in the low-field: Leveraging the power of machine learning}
\author[1]{Andreas Kofler}
\author[2]{Dongyue Si}
\author[1]{David Schote}
\author[2,3]{Rene M Botnar}
\author[1]{Christoph Kolbitsch}
\author[2,3]{Claudia Prieto}
\affil[1]{Physikalisch-Technische Bundesanstalt (PTB), Braunschweig and Berlin, Germany}
\affil[2]{King's College London, London, United Kingdom}
\affil[3]{Pontificia Universidad Católica de Chile, Santiago de Chile, Chile}
\begin{document}

\maketitle

\abstract{}
Recent innovations in Magnetic Resonance Imaging (MRI) hardware and software have reignited interest in low-field ($<1\,\mathrm{T}$) and ultra-low-field MRI ($<0.1\,\mathrm{T}$). These technologies offer advantages such as lower power consumption, reduced specific absorption rate, reduced field-inhomogeneities, and cost-effectiveness, presenting a promising alternative for resource-limited and point-of-care settings. However, low-field MRI faces inherent challenges like reduced signal-to-noise ratio and therefore, potentially lower spatial resolution or longer scan times.

This chapter examines the challenges and opportunities of low-field and ultra-low-field MRI, with a focus on the role of machine learning (ML) in overcoming these limitations. We provide an overview of deep neural networks and their application in enhancing low-field and ultra-low-field MRI performance. Specific ML-based solutions, including advanced image reconstruction, denoising, and super-resolution algorithms, are discussed. The chapter concludes by exploring how integrating ML with low-field MRI could expand its clinical applications and improve accessibility, potentially revolutionizing its use in diverse healthcare settings.


\section{Introduction}\label{sec:introduction}

Magnetic Resonance Imaging (MRI) is an essential tool for the early detection, risk stratification, prognosis, treatment selection, and monitoring of many diseases, including cancer, cardiovascular disease, metabolic, musculoskeletal, and brain disorders, among many others. Its ability to produce multi-contrast and multi-parametric images of soft tissues, coupled with its non-invasive and radiation-free nature, makes it a highly valuable tool in clinical practice.

Over the past five decades, the technology behind MRI has undergone significant advancements, especially in terms of the magnetic field strengths used for imaging. Early MRI systems operated at low field strengths ($0.15\,\mathrm{T}$ to $0.35\,\mathrm{T}$) \cite{Hori2021,Sarracanie2020,Hennig2023}, and while they offered important diagnostic insights, they were limited by low signal-to-noise ratio (SNR) and image resolution. Over time, several advancements led to the development of systems operating at higher field strengths, such as $1.5\,\mathrm{T}\,$ and $3\,\mathrm{T}$, which are now considered the clinical standard due to their superior SNR and image quality \cite{Niendorf2017,Dumoulin2018}. Recent developments have even pushed field strengths to ultra-high levels ($\geq 3\,\mathrm{T}$), including $5\,\mathrm{T}$, $7\,\mathrm{T}$ and beyond, further enhancing the spatial and temporal resolution of MRI \cite{Niendorf2017,Wei2023,Reiter2021}. However, high-field MRI has its challenges \cite{Winter2015}. While it offers improved imaging quality, higher fields are associated with certain drawbacks, including increased costs, higher specific absorption rates (SAR), and higher radiofrequency-induced heating, which are particularly problematic for patients with implants. Although MR-safe implants exist, they typically lead to more severe image artefacts at higher fields, see e.g.\ \cite{van2021characterization}. However, advances in state-of-the-art MRI are driven not solely by the higher magnetic field strength but also by a wide range of other remarkable developments in hardware and software, including improved coil arrays, digital receiver technology, and reconstruction algorithms, among many others. In recent years, there has been renewed interest in low-field MRI ($<1\,\mathrm{T}$) and ultra-low-field MRI ($<0.1\,\mathrm{T}$), taking full advantage of these cutting-edge developments. With lower power consumption, reduced SAR, decreased field inhomogeneity, and lower installation and maintenance costs, low-field MRI presents an opportunity to make MRI more accessible and cost-effective, particularly in resource-limited settings. Furthermore, ultra-low-field MRI offers the potential for small, mobile systems without special installation requirements, opening the door to bringing this technology to point-of-care settings, such as emergency rooms and general practitioner practices.

Despite its advantages, low-field and ultra-low-field MRI faces several technical challenges, particularly in terms of reduced SNR, spatial resolution, and scan times compared to high-field systems. These limitations could restrict its broader adoption and application, especially in cases where high-resolution and high-sensitivity imaging is required. Recently, several machine learning (ML) techniques have been proposed that can aid in overcoming these limitations and enhance the performance of low-field MRI. ML-based undersampled image reconstruction techniques, denoising methods, and super-resolution algorithms offer the potential to compensate for the reduced SNR inherent in low-field systems. Moreover, ML can assist in automating image analysis and optimizing acquisition protocols, as well as in image post-processing and analysis, thereby improving the diagnostic utility and efficiency of low-field MRI. 

In this chapter, we explore the challenges associated with low-field and ultra-low-field MRI and highlight the growing role of ML in addressing these limitations. A brief introduction to deep neural networks is provided to give context for the discussed methods.  A summary of ML-based solutions for image reconstruction acquisition and post-processing in low-field MRI is also included. Finally, we discuss the potential of integrating ML-based solutions with low-field MRI to broaden its use in clinical practice.

\section{Challenges of low-field MRI}\label{sec:history}

In recent years, contemporary low-field ($< 1\,\mathrm{T}$) and ultra-low-field ($< 0.1\,\mathrm{T}$) MRI has been shown to provide potential solutions to common historical limitations of lower field MRI, with several advantages including low power, low SAR, less field inhomogeneity and lower costs \cite{Simonetti2017,Gilk2023,Marques2019,Arnold2023,Shetty2023,Liu2021,Zhao2024}. Furthermore, lower-field MRI has advantages in comparison to high-field strengths for some applications including MRI-guided interventions, imaging anatomy near air-tissue interfaces, and efficient image acquisition methods such as spiral out acquisition \cite{Rogers2023,Atalay2001,Restivo2020}. An early attempt at contemporary low-field MRI was conducted by modifying a $1.5\,\mathrm{T}$ scanner to operate at $0.55\,\mathrm{T}$ \cite{Campbell-Washburn2019}. This study demonstrated that standard pulse sequences can provide adequate image quality in routine imaging on the high-performance ramped-down $0.55\,\mathrm{T}$ system in comparison with $1.5\,\mathrm{T}$. Commercial low-field systems such as $0.5\,\mathrm{T}$ Synaptive Evry intraoperative scanner, $0.55\,\mathrm{T}$ Siemens Magnetom Free.Max general-purpose scanners and ultra-low-field systems such as $64\,\mathrm{mT}$ Hyperfine's Swoop portable MRI system have also been introduced subsequently. With lower costs, they have the potential to increase the adoption of MRI over the world. In addition to commercial scanners, there are several research groups and consortia developing ultra-low-field prototype systems for brain- or extremities-only imaging or whole-body imaging, which can be portable and more patient-friendly, enabling point-of-care applications \cite{Liu2021,Zhao2024}. Although low-field MRI has many advantages in comparison to high-field systems and can bring new applications to the clinic, it still faces many challenges \cite{Arnold2023,Shetty2023}. 

\subsection{Signal-to-Noise Ratio (SNR)}
\begin{figure}
    \centering
    \includegraphics[width=0.95\linewidth]{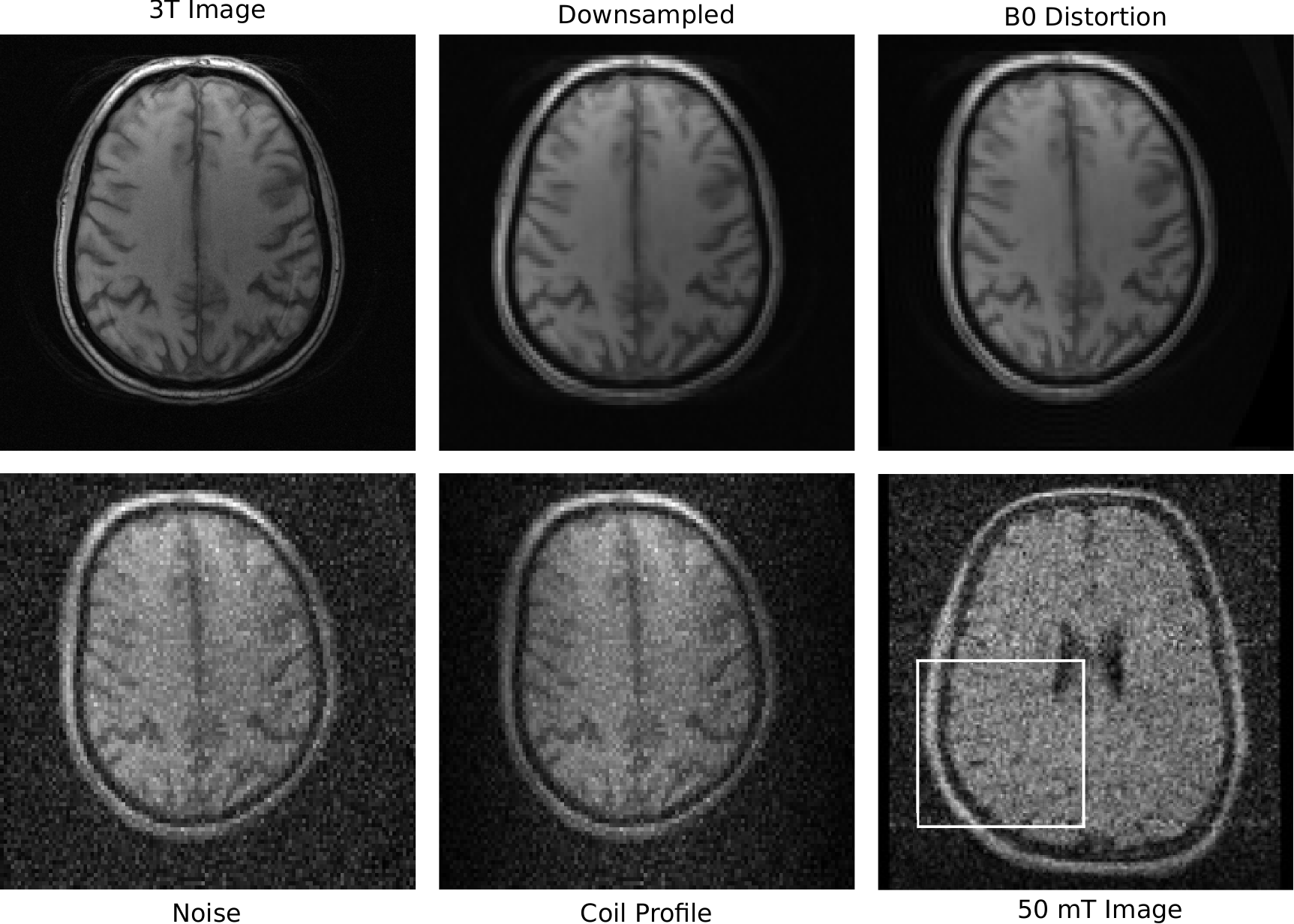}
    \caption{Comparison of a high-field ($3\,\mathrm{T}$) and a corresponding simulated ultra-low-field (50mT) MR image. Lower field systems typically only achieve a decreased spatial resolution (\textit{downsampled}), are exposed to strong B0-inhomogeneities (\textit{B0 distortion}), poor SNR (\textit{noise}) and use receiver coils with high Q-factors (\textit{coil profile}). The upper left image is an example taken from the fastMRI brain dataset \cite{zbontar2018fastmri}, the intermediate images were obtained by applying the just mentioned different degradation steps. The lower right image is taken from \cite{de2022deep} and was acquired with the $50\,mT$ ultra low-field scanner described in \cite{o2021vivo}. Image courtesy of Andrew Webb.}
    \label{fig:high_vs_low_field}
\end{figure}

SNR is a critical factor in determining image quality in MRI, especially in clinical settings where the visualization of fine anatomical details is vital for accurate diagnosis. The main drawback of lower field imaging is the inherent reduction in SNR due to the weaker magnetic field, see Figure \ref{fig:high_vs_low_field}, as the magnetization signal is proportional to the B0 field strength \cite{Marques2019}. At lower magnetic fields, the net magnetization of protons in the body is reduced, leading to a weaker signal. This diminished signal must compete with the noise inherent to the systems, which remains relatively comparable regardless of field strength, assuming the same acquisition parameters and hardware \cite{Macovski1996}. As a result, the overall SNR decreases, making it more challenging to produce high-quality images. The loss of SNR at low-field can be partly compensated by using different acquisition sequences or modifying imaging parameters such as bandwidth and flip angle \cite{Restivo2020,https://doi.org/10.1002/jmri.24163}. Nevertheless, these adjustments can introduce changes in contrast and potential artefacts that must be carefully considered. 

High spatial resolution is critical in many clinical scenarios, such as in neuroimaging, where fine details of brain structures must be discerned. The lower SNR (per unit time) and lower scan efficiency of lower field systems often translate into lower achievable spatial resolution. Low-resolution MRI could be sufficient for broad assessments and follow-up scans. However, to enable the utility in cases requiring detailed high-resolution imaging, specific optimization may be required for low-field especially ultra-low-field MRI to match the image quality and diagnostic ability of higher-field MRI systems \cite{Deoni2021}.

Figure \ref{fig:high_vs_low_field} shows an exemplary comparison of a brain MR image obtained from a $3\,\mathrm{T}$ scanner and its potential counterpart obtained by simulating the different degradation processes that are inherently linked to a low-field image reconstruction problem. 

\subsection{Scan Time}
Scan time is an important practical factor in clinical applications. However, low-field MRI usually exhibits longer scan times. On the one hand, as the SNR decreases at low field, the SNR per unit time is also reduced. Therefore, to achieve sufficient SNR, low-field imaging usually requires more time compared to higher field strengths. For example, using a lower acquisition bandwidth can improve the SNR, but at the same increases the repetition time and thus reduces the scan time efficiency. Additionally, $k$-space undersampling at low-field may achieve lower acceleration factors, and repeating scan and averaging signals may also be necessary to obtain high-SNR images \cite{Varghese2023}.\\
On the other hand, the efficiency of MRI is also influenced by the performance of gradient and radiofrequency (RF) systems \cite{Marques2019,Roemer1990}. Gradients are used for spatially encoding the MRI signal. Less powerful gradients lead to longer repetition and echo times, thus lower scan efficiency. Similarly, the RF transmitter and receiver system, responsible for exciting the protons and receiving the emitted signals, also influence the duration of RF pulses and repetition time. In low-field MRI, these systems often face significant limitations. Commercial low-field MRI scanners, often come with more basic gradient and RF systems. These limitations can result in longer acquisition times, making the scanning process less efficient and more prone to motion artefacts. 

\subsection{Relaxation Differences}
Relaxation times (T1 and T2) are fundamental to the contrast mechanisms in MRI, influencing how different tissues appear in the resulting images. However, as can be seen in Table \ref{tab:Table_T1T2}, these relaxation times vary with field strength, leading to different image contrast at low-field compared to high-field systems \cite{Rooney2007,Peters2007}. T1 relaxation times tend to decrease at lower fields, resulting in reduced contrast between tissues that would normally appear distinct in high-field imaging \cite{Rooney2007}.
Chemical shift, the difference in resonance frequencies of different molecular environments, is also reduced at lower field strengths \cite{Delfaut1999}. This may make the differentiation between fat and water in tissues difficult, potentially affecting the accuracy of diagnoses in areas such as liver imaging or breast MRI, where fat suppression techniques are commonly used \cite{Tian2024}.
In addition, the efficacy of gadolinium-based contrast agents is field-dependent. The relaxivity of gadolinium is reduced at lower magnetic fields especially for ultra-low-field, diminishing its contrast-enhancing effects \cite{Arnold2023}. Thus, further experiments are required to validate the utility of gadolinium-based contrast-enhanced MRI studies in low-field systems, which may require higher contrast dose or the use of new contrast agents with higher relaxivity \cite{Waddington2024}. 

\begin{table}[h!]
\centering
\renewcommand{\arraystretch}{1.2}
\footnotesize
\begin{tabular}{c|c|c|c|c}
\hline
\textbf{Tissue} & \textbf{T1 (msec)} & \textbf{T2 (msec)} & \textbf{T2* (msec)} & \textbf{T2*-to-T2 Ratio (\%)} \\ \hline
\rowcolor[HTML]{E0E0E0}
\textbf{White matter} & & & & \\
$0.55\,\mathrm{T}$ & 493 $\pm$ 33 & 89 $\pm$ 9 & 72 $\pm$ 12 & 81 \\
$1.5\,\mathrm{T}$ & 608--884 & 54--96 & 48--60 & 72 \\ \hline
\rowcolor[HTML]{E0E0E0}
\textbf{Gray matter} & & & & \\
$0.55\,\mathrm{T}$ & 717 $\pm$ 82 & 112 $\pm$ 7 & 86 $\pm$ 9 & 77 \\
$1.5\,\mathrm{T}$ & 1002--1304 & 93--109 & 67--79 & 72 \\ \hline
\rowcolor[HTML]{E0E0E0}
\textbf{Myocardium} & & & & \\
$0.55\,\mathrm{T}$ & 701 $\pm$ 24 & 58 $\pm$ 6 & 47 $\pm$ 4 & 80 \\
$1.5\,\mathrm{T}$ & 950--1030 & 40--58 & 30--37 & 68 \\ \hline
\rowcolor[HTML]{E0E0E0}
\textbf{Arterial blood} & & & & \\
$0.55\,\mathrm{T}$ & 1122 $\pm$ 85 & 263 $\pm$ 27 & \dots & \dots \\
$1.5\,\mathrm{T}$ & 1441--1898 & 254--290 & \dots & \dots \\ \hline
\rowcolor[HTML]{E0E0E0}
\textbf{Liver} & & & & \\
$0.55\,\mathrm{T}$ & 339 $\pm$ 31 & 66 $\pm$ 6 & 43 $\pm$ 4 & 65 \\
$1.5\,\mathrm{T}$ & 576--586 & 46--55 & 26--33 & 59 \\ \hline
\rowcolor[HTML]{E0E0E0}
\textbf{Lung} & & & & \\
$0.55\,\mathrm{T}$ & 971 $\pm$ 62 & 61 $\pm$ 11 & 10 $\pm$ 2 & 17 \\
$1.5\,\mathrm{T}$ & 1171--1333 & 41 & 1--2 & 5 \\ \hline
\rowcolor[HTML]{E0E0E0}
\textbf{Kidney cortex} & & & & \\
$0.55\,\mathrm{T}$ & 651 $\pm$ 48 & 101 $\pm$ 7 & 82 $\pm$ 17 & 82 \\
$1.5\,\mathrm{T}$ & 690--966 & 65--87 & 49--51 & 70 \\ \hline
\rowcolor[HTML]{E0E0E0}
\textbf{Fat} & & & & \\
$0.55\,\mathrm{T}$ & 187 $\pm$ 10 & 93 $\pm$ 16 & \dots & \dots \\
$1.5\,\mathrm{T}$ & 288--343 & 53--84 & \dots & \dots \\ \hline
\end{tabular}
\caption{A summary of tissue relaxation parameters at $0.55\,\mathrm{T}$ compared to their reference $1.5\,\mathrm{T}$ Values. Table reproduced from \cite{Campbell-Washburn2019}. Note that the values reported for the $0.55\,\mathrm{T}$ were obtained from the experiments performed by the authors in \cite{Campbell-Washburn2019} and are therefore given in terms of mean$\pm$standard deviation. In contrast, the values for $1.5\,\mathrm{T}$ are taken from different literature sources. We refer the reader to \cite{Campbell-Washburn2019} for more details.}\label{tab:Table_T1T2}
\end{table}

\subsection{Advanced Imaging Techniques}
Advanced imaging techniques such as functional MRI (fMRI), diffusion tensor imaging (DTI), and MR spectroscopy require higher field strengths to achieve the SNR and sensitivity needed for meaningful clinical and research applications \cite{Dumoulin2018, Tournier2019, Rodgers2014}. These techniques rely on the subtle differences in signal that are more readily detected at higher fields. For example, fMRI, which maps brain activity by detecting changes in blood oxygenation, requires high spatial and temporal resolution to accurately localize brain functions. DTI, used to map white matter tracts in the brain, requires high gradient strength and SNR to track the diffusion of water molecules along nerve fibers \cite{Tournier2019}. Similarly, MR spectroscopy, which analyzes the chemical composition of tissues, requires a strong signal to detect the low-concentration metabolites of interest \cite{Rodgers2014}.
Low-field MRI systems can struggle to achieve the necessary resolution and sensitivity for these advanced techniques, which may limit their application in both clinical practice and research \cite{Bhat2021}. 

\subsection{Low-Field MRI in a High-Field World}
As high-field MRI has become the standard in clinical practice, regulatory bodies and industry standards have been developed with these systems in mind \cite{Hennig2023}. Low-field MRI, by contrast, often falls outside these established norms, presenting challenges for its broader adoption and integration into healthcare systems. The deployment of low-field MRI systems may require different standards for safety, performance, and image quality assessment. For example, the specific absorption rate (SAR) differs between high- and low-field systems, necessitating different safety guidelines \cite{Gilk2023}. Additionally, regulatory approval processes for low-field MRI systems and their associated hardware and software may be more complex, potentially slowing their adoption. Moreover, the lack of standardized protocols for low-field MRI can lead to variability in image quality and diagnostic accuracy between different systems and institutions. This variability poses a challenge for radiologists, who must interpret images that may not conform to the high standards set by their high-field counterparts. As a result, there is a pressing need for the development of tailored guidelines and standards for low-field MRI to ensure consistent and reliable imaging outcomes \cite{Islam2023}.

\section{Opportunities for AI-based solutions to advance low-field MRI}\label{sec:opportunities}
Machine Learning- (ML) and Artificial intelligence (AI)-based solutions promise to address the inherent limitations of low-field MRI and unlock its potential for broader, more accessible clinical use, by exploiting algorithms developed to enhance image quality and improve diagnostic accuracy \cite{Arnold2023}. These algorithms are designed to compensate for the lower SNR, reduced spatial resolution, and other challenges specific to low-field imaging \cite{LopezSchmidt2023}. For example, AI-driven reconstruction algorithms can optimize the use of available data, reducing the need for long scan times and improving the overall efficiency of low-field MRI \cite{Zhu2018}. Deep learning models, can significantly improve the quality of low-field MR images by enhancing resolution (with super-resolution techniques), denoising, and correcting for artifacts, making them more comparable to higher-field images \cite{DeLeeuwdenBouter2022,Le2021}. These techniques can also assist in the automatic post-processing, such as segmentation of tissue, improving the accuracy and efficiency of image analysis \cite{Bernard2018}. This is crucial in settings where radiologist expertise may be limited or unavailable, broadening access to MRI diagnostics. AI-based approaches have also been proposed to eliminate electromagnetic interference and enable RF-shielding-free ultra-low-field MRI scanners \cite{Liu2021,Zhao2024}. Furthermore, by leveraging AI to optimize pulse sequences specifically for low-field environments, it is possible to maximize image quality and diagnostic value, tailoring protocols to the constraints of these systems. By leveraging the power of AI, low-field MRI systems can produce images that are closer in quality to those obtained with high-field systems, thereby expanding their utility in clinical practice. As AI technology continues to evolve, it holds the potential to significantly mitigate the challenges of low-field MRI, paving the way for broader adoption and application of these systems.

Different applications of AI-based solutions to advance low-field MRI are discussed in the following. Before discussing these methods in detail, we provide the reader with a brief introduction to basic concepts and terms of neural networks (NNs) and deep learning (DL) that are needed to understand some aspects of the methods discussed in this Chapter. A more detailed description of these concepts can be found in Chapter 2. 

\subsection{Deep Neural Networks in a Nutshell}

Deep neural networks are functions $f_{\theta}$ between two different spaces $V$ and $W$ that are parameterized by a set of parameters $\theta $ (typically $\theta \in \R^\ell$, although networks with complex-valued parameters exist as well) that one wants to learn based on some data. Thereby, the function $f_{\theta}:V \rightarrow W, v \mapsto w:=f_{\theta}(v)$ consists of a composition of multiple functions (hence the term \textit{deep} learning) of the form $f_{\theta_i}$ for $i=1,\ldots, N_{\mathrm{layers}}$ that have the form $f_{\theta_i} = \phi_i(\mathbf{W}_i \cdot + \mathbf{b}_i)$ for appropriately sized $\mathbf{W}_i$ and $\mathbf{b}_i$. The variables $\mathbf{W}_i$ and $\mathbf{b}_i$ are typically referred to as the weights and biases of the $i$-th layer, respectively, while the function $\phi_i$ is typically a non-linear activation function that is applied component-wise to the resulting vector. Then, a (feed-forward) neural network $f_{\theta}$ with trainable parameters $\theta$ can be constructed as the composition of the respective layers
\begin{equation}\label{eq:neural_network}
    f_{\theta} := \bigcircle_{i=1}^{N_{\mathrm{layers}}} f_{\theta_i}, \quad \text{ with } \quad \theta:=\bigcup_{i=1}^{N_{\mathrm{layers}}} \theta_i,
\end{equation}
A large variety of different neural network architectures exists nowadays. Because the presentation of various neural network architectures is out of scope for this chapter, here we only mention the U-Net \cite{ronneberger2015u} and refer the reader to Chapter 2 for other architectures. The U-Net features a symmetric "U" shape, consisting of a contracting path (encoder) that captures context by downsampling and an expanding path (decoder) that enables feature localization through upsampling. Skip connections between the corresponding layers of the encoder and decoder paths allow for the preservation of high-resolution features. The U-Net is currently, without a doubt, one of the most widely used architectures and ubiquitous in the field of deep learning-based image processing. 

\subsection{Network Training}\label{subsec:network_training}
Training a deep neural network $f_{\theta}$ refers to the process of tuning the parameters $\theta$ on a dataset $\mathcal{D}$. This is typically pursued to later employ $f_{\theta}$ as a model for prediction, i.e.\ after having obtained a suitable set of parameters $\theta$, one is interested in being able to predict new meaningful outputs using the learned model $f_{\theta}$.

Assuming the existence of a dataset of input-target pairs $\mathcal{D}:= \{ (v_i, w_i)_{i=1}^{N_{\mathrm{samples}}}\}$,
training a neural network $f_{\theta}$ is typically carried out employing gradient descent-like methods, i.e.\ one constructs a sequence of parameters $\{\theta_k\}_{k \in \mathbb{N}}$ by
\begin{equation}\label{eq:gradient_descent}
    \theta_{k+1} := \theta_k - \tau_k (\nabla_{\theta}\mathcal{L})(\theta_k),
\end{equation}
where $\mathcal{L}$ represents a so-called loss function and $\tau_k$ a (possibly iteration-dependent) step-size. The loss function $\mathcal{L}$ typically takes the form 
\begin{equation}\label{eq:loss_function}
    \mathcal{L}(\theta):= \frac{1}{|\mathcal{D}|} \sum_{(v,w) \in \mathcal{D}} l\big(f_{\theta}(v), w) + r(\theta),
\end{equation}
where $l(\, \cdot \, , \, \cdot \,)$ is an appropriately chosen similarity/discrepancy measure that is used to measure how close the predicted output is to its corresponding target, and $r$ is some regularization term that can be further employed to avoid overfitting to the data, i.e.\ to avoid that the network is not able to generalize to unseen examples.

As the set of parameters $\theta$ is typically rather large, training a neural network is carried out by the following procedure. One splits the available dataset in three disjoint subsets $\mathcal{D}_{\mathrm{train}}, \mathcal{D}_{\mathrm{validation}}$ and $\mathcal{D}_{\mathrm{test}}$ and attempts to approximately minimize Eq.\,\eqref{eq:loss_function} over the training set $\mathcal{D}_{\mathrm{train}}$, while at the same time monitoring the value of the loss-function $\mathcal{L}$ when evaluated over the validation set. The latter is an important step that is used to ensure that the model generalizes. The underlying idea is that the performance of the model when measured with respect to the chosen metric $l$ over the validation set can be used as a proxy for the (in practice) unknown performance of the model at prediction time. The validation set is typically also used to choose different hyper-parameters of the neural network $f_{\theta}$, e.g.\ the number of layers or the size of the weights and biases, or hyper-parameters of the learning procedure, e.g.\ the specific optimization routine, the learning rate or the number of gradient-descent steps. 

Note that, for computational reasons, in Eq.\,\eqref{eq:gradient_descent}, one often does not compute the exact gradient of the loss function in Eq.\,\eqref{eq:loss_function}, as it involves the summation over the entire dataset, but rather approximates the loss function using a small number of (randomly chosen) samples at each iteration. This approach is referred to as stochastic gradient descent and further introduces some form of implicit regularization.

Further, note that, instead of employing plain gradient descent techniques as mentioned in  Eq.\,\eqref{eq:gradient_descent}, more sophisticated training algorithms are nowadays used to train modern architectures, e.g.\ RMSprop \cite{hinton2012neural}, AdaDelta \cite{zeiler2012adadelta} or Adam \cite{kingma2014adam}, to name a few. See  \cite{ruder2016overview} for an exhaustive review.

Last, we note that the terms \textit{validation} and \textit{test} datasets are sometimes interchangeably used but their exact meaning typically emerges from the context. Here, we always refer to the test data as the dataset that is used to evaluate the performance of the final model for some metric.

Next, we give an overview of different AI methods that have been applied or can potentially be applied to low-field MRI to enhance image quality and diagnosis. We start with methods that can be used to adapt the acquisition tasks such as automatic planning of imaging planes that could potentially be applied for low-field MRI. We then proceed with image reconstruction methods and discuss the basic properties of deep learning methods to accelerate the acquisition and/or enhance image quality. We categorize the different deep learning-based image reconstruction approaches in 1) post-processing reconstruction techniques, which are applied to an initial (low quality) image estimate, 2) model-unaware techniques, which do not take known information about the image process into account and 3) physics-informed approaches, which in contrast do include this information. As the vast majority of deep learning image reconstruction methods have been developed for high-field MRI, some of these methods are also mentioned here as they are needed to understand the current research landscape. Further details about image reconstruction can be found in Part III of this book, Chapters 7 to 11.  We then finally briefly discuss deep learning approaches for post-processing analysis tasks (such as segmentation, image quality control, radiomics, and classification).  Further description of AI-based image post-processing can be found in Section IV, Chapters 12-15.

\subsection{AI solutions for image planning and acquisition in low-field MRI}
One main advantage of MRI compared to most other medical imaging techniques is the ability to adapt the image orientation to the organ of interest. This can strongly improve the diagnostic quality and help with standardized diagnosis. Depending on the organ of interest, the planning of the correct image orientation can be very time-consuming. Significant training is also required to ensure accurate and reproducible scan orientations.

One particularly challenging application is fetal imaging. In \cite{silva2024mrm}, the authors presented an approach for automatic scan planning for fetal brain imaging at $0.55\,\mathrm{T}$ using a 3D U-Net architecture. The brain of the fetus is automatically detected in the womb of the mother and a set of landmarks (e.g. eyes) are identified. Based on these landmarks, scan planning is carried out. This approach reduced scan planning from several minutes down to a few seconds. Training was carried out on high-field data. A similar approach was also used to allow for automated scan planning for fetal flow imaging of the aorta and the umbilical vein\cite{silva2024arxiv}. 

An important advantage of ultra-low-field MR devices is their portability. Rather than bringing patients to a dedicated MR scanner room, the MR scanner can travel to the patient. This offers many advantages and new possibilities for healthcare. One challenge is though, that the MR scans are not carried out in a controlled environment anymore. Electromagnetic noise, in particular, can severely impair the quality of images. Approaches that measure electromagnetic noise using dedicated coils and then utilize CNN or U-Net architectures to predict the noise contribution to the acquired data have been presented \cite{Su2022tmi, Zhao2024mrm}. Denoising is then a rather simple step of subtracting the estimated noise contribution prior to image reconstruction. 

In AUTOSEQ \cite{zhu2018automated}, although not specifically designed and applied to low-field MRI, the authors explore a method for the automatic development of new non-trivial pulse sequences for the data acquisition in MRI. They achieve this by employing a reinforcement learning framework and recasting the problem of pulse sequence development as a game of perfect information that an AI agent learns to play. As initial results, the authors reported that their agent, apart from being able to learn a canonical gradient echo pulse sequence, also succeeded in learning to generate more non-intuitive pulse sequences.
Due to its general nature, the approach presented in \cite{zhu2018automated} holds the promise to be applicable to low-field MRI as well.

\subsection{AI solutions for Image Reconstruction in low-field MRI}

Here, we first briefly formalize the image reconstruction problem of interest and then present different machine learning-based approaches that can be employed to obtain a meaningful approximation of the unknown sought image. 

\subsubsection{Forward Problem Formulation}
A discretized version of the MR image acquisition (i.e.\ the forward problem) can be expressed as 
\begin{equation}\label{eq:forward_problem}
    \sd = \Ed \boldsymbol{\rho}_{\mathrm{true}} + \boldsymbol{\eta},
\end{equation}
where $\boldsymbol{\rho}_{\mathrm{true}} \in \C^N$ denotes the vector representation of the (unknown) complex-valued image with $N$ being the total number of voxels, e.g.\ $N=N_x\cdot N_y$ for a 2D image or $N=N_x\cdot N_y\cdot N_z$ for a 3D volume. The forward operator $\Ed$ models the data acquisition and involves several quantities such as coil sensitivity maps, the set of $k$-space trajectories along which the image is sampled, field maps inhomogeneities and possibly other relaxation effects as well as other set up- and sequence-specific parameters. Thus, note that, in general, the operator $\Ed$ in Eq.\,\eqref{eq:forward_problem} is non-linear, but the vast majority of the literature considers an idealized set-up to be able to work with a linear acquisition model. The vector $\boldsymbol{\eta}\in \C^K \sim \mathcal{N}(\boldsymbol{\mu},  \boldsymbol{\Sigma})$ with $K$ being the total number of measured $k$-space points denotes complex-valued Gaussian noise with mean vector $\boldsymbol{\mu}$ and covariance matrix $\boldsymbol{\Sigma}$ (which is often modeled by $\sigma^2 \Id$ with variance $\sigma^2$) and $\sd \in \C^M$ denotes the acquired $k$-space data. 

For example, in a multi-coil setting, when $N_{\mathrm{c}}$ receiver coils are used, the forward operator is often modeled by
\begin{eqnarray}
    &\Ed&:=(\Id_{N_{\mathrm{c}}} \otimes \Fd) \Cd, \label{eq:forward_operator_multicoil_A}  \\
    &\Cd&:=[\Cd_1,\ldots,\Cd_{N_{\mathrm{c}}}]^\trans, \quad \text{ with } \quad \Cd^\herm \Cd = \Id_N\label{eq:forward_operator_multicoil_C} \\
    &\Cd_{j}&:=\mathrm{diag}(\mathbf{c}_j), \quad \mathbf{c}_j \in \C^{N\times N}, j=1,\ldots, N_{\mathrm{c}},
\end{eqnarray}
where $\Id_{N_{\mathrm{c}}}$ denotes the $N_{\mathrm{c}}\times N_{\mathrm{c}}$ identity matrix, $\Fd$ is a Fourier-encoding operator and  $\Cd_{j}$ represents the $j$-th coil sensitivity map. For single-coil acquisitions, $N_{\mathrm{c}}=1$ and $\Cd$ in Eq.\,\eqref{eq:forward_operator_multicoil_C} reduces to the identity matrix.

Further, in so-called super-resolution applications \cite{de2022deep, iglesias2022quantitative, iglesias2022accurate, laguna2022super, lau2023pushing, donnay2024super} (see also Chapter 8), the forward problem in Eq.\,\eqref{eq:forward_problem} is extended by 
\begin{equation}\label{eq:forward_operator_super_resolution}
     \sd = \Ed \Rd \boldsymbol{\rho}_{\mathrm{true}} + \boldsymbol{\eta}, 
\end{equation}
with the understanding that $\boldsymbol{\rho}_{\mathrm{true}}$ denotes the high-resolution image that is sub-sampled by a low-resolution operator $\Rd:\C^N \rightarrow \C^{N^\prime}$ and whose $k$-space data is acquired. The goal is then to recover the high-resolution image from the $k$-space data of the low-resolution image. 

Further, since in MRI, data acquisition is a time-consuming process, it is sometimes desirable to accelerate the measurement by only acquiring a portion of the $k$-space that would be required to be sampled. In this case, the forward problem is given as in Eq.\,\eqref{eq:forward_problem}, but we emphasize the acquisition of only a sub-portion of the required $k$-space data by writing $\Ed_J$ instead of $\Ed$ and $\sd_J$ instead of $\sd$ with the understanding that $\Ed_J$ samples $k$-space data only at locations indexed by an index set $J$, i.e.\ the Fourier-encoding operator in Eq.\,\eqref{eq:forward_operator_multicoil_A} is analogously denoted by $\Fd_J$ instead of $\Fd$.

Regardless of the operator $\Ed$ being linear or non-linear, the reconstruction of the image $\boldsymbol{\rho}_{\mathrm{true}}$ from the observed $k$-space data $\sd$ often represents an ill-posed problem. The number of measured data points $K$ and its relation to the number of sought voxels $N$ implicitly defines the nature of the ill-posedness. For $K<N$, e.g.\ for accelerated undersampled acquisitions or super-resolution applications, problem Eq.\,\eqref{eq:forward_problem} is under-determined, meaning that even in the linear case, an infinite number of solutions exists. Since $N^{\prime}< N$, the super-resolution problem in Eq.\eqref{eq:forward_operator_super_resolution} is clearly ill-posed by being under-determined.

Even for $K>N$, problem Eq.\,\eqref{eq:forward_problem} is ill-conditioned for non-Cartesian sampling trajectories.

For the considered case of low-field MRI, an additional problem is that due to the lower field strength of the static magnetic field B0, the signal generated by the transverse magnetization of the object to be imaged is inherently lower than when measured at higher field strengths. This results in a systematically lower SNR, which can be modeled by the relationship of the mean of $\boldsymbol{\rho}_{\mathrm{true}}$ and the noise variance $\sigma^2$ in Eq.\,\eqref{eq:forward_problem}, and poses a further challenge to the image reconstruction process.

A widely followed approach to still be able to reconstruct  images useful for clinical diagnosis is to formulate a variational problem of the type
\begin{equation}
    \underset{\boldsymbol{\rho}}{\min} \;  \frac{1}{2}\|\Ed \boldsymbol{\rho} - \sd\|_2^2 + \mathcal{R}(\boldsymbol{\rho}),\label{eq:variational_problem}
\end{equation}
where the first term in Eq.\,\eqref{eq:variational_problem} tackles the data-consistency of the sought solution and $\mathcal{R}(\boldsymbol{\rho})$ is a regularization term that is used to impose desirable properties on the image. Note that, although in principle different data-discrepancy measures can be employed as well, the use of the squared $\ell_2$-norm is motivated by the fact that Gaussian noise is the dominant noise present in the MRI acquisition \cite{brown2014magnetic}.

The study of suitable regularization functionals has been an active field of research in the last decades. Deep learning-based methods that implicitly learn features from data have also been proposed as discussed for low-field imaging in the following and more generally in Chapter 7. 

\subsubsection{Supervised vs Self-Supervised Training for Image Reconstruction}

In Subsection \ref{subsec:network_training}, we have implicitly assumed the existence of a dataset consisting of input-target pairs, e.g.\ $\mathcal{D}:=\{(\sd^{(i)}, \boldsymbol{\rho}_{\mathrm{true}}^{(i)}) |\, \sd^{(i)} = \Ed \boldsymbol{\rho}_{\mathrm{true}}^{(i)} + \boldsymbol{\eta}^{(i)}\}$, which means the network can be trained by adjusting the parameters $\theta$ such that $f_{\theta}(\sd^{(i)})\approx \boldsymbol{\rho}_{\mathrm{true}}^{(i)}$ for each pair. For example, by setting the discrepancy function $l$ in Eq.\,\eqref{eq:loss_function} to $l(\,\cdot, \, \cdot\,) = \|\, \cdot\, - \,\cdot\, \|_2^2$, one can  minimize the supervised loss-function
\begin{equation}\label{eq:loss_function_supervised}
    \mathcal{L}^{\mathrm{sup}}(\theta):= \frac{1}{|\mathcal{D}|} \sum_{(\sd, \boldsymbol{\rho}_{\mathrm{true}}) \in \mathcal{D}}  \|f_{\theta}(\sd) - \boldsymbol{\rho}_{\mathrm{true}} \|_2^2.
\end{equation}

However, deep neural networks are known to require a rather large dataset to be properly trained. In the context of image reconstruction, this corresponds to having access to many target images of good quality, which can be costly, challenging, or, in some cases, even impossible to obtain, which is especially the case for low-field MRI.

This issue is often circumvented by relying on simulations and the retrospective data generation of input-target pairs, see e.g.\ \cite{de2022deep, schote2024joint, iglesias2022accurate},  but can possibly yield a performance gap in the obtained model, especially when the distribution shifts from the training to the test data is rather large \cite{darestani2022test}. Further, to enrich the set of target image data used for training, it is usually advised to employ data augmentation techniques such as simple image rotations, image flipping, contrast changes, or more involved techniques, e.g.\ elastic deformations \cite{schlemper2017deep}.  

Self-supervised training strategies as the ones discussed next can overcome these challenges by exploiting the structure of the considered problem. Recall that for image reconstruction tasks, our available input data (e.g. acquired $k$-space data) is in general generated according to a known forward model as in Eq.\,\eqref{eq:forward_problem}. Rather than comparing the output of the network to a known target, we can apply the forward model to the network output and compare the obtained result to the available input data. Although intuitive, this approach can suffer from overfitting problems because the employed neural networks are typically overparameterized. Thus, a mechanism that prevents the learning procedure from yielding sub-optimal parameters $\theta$ is required. Strategies to avoid overfitting are typically based either on early stopping techniques \cite{ulyanov2018deep}, on restricting the family of mappings that can be learned by imposing certain constraints on the network architectures, i,e.\ on the network's topology \cite{heckel_deep_2018}, invertibility of the obtained mapping  \cite{etmann2020iunets}, employing regularization on the weights, e.g.\ dropout \cite{srivastava2014dropout}, batch-normalization \cite{ioffe2015batch}, spectral-normalization \cite{miyato2018spectral}, to only name a few, or by employing appropriate loss functions \cite{lehtinen2018noise2noise, moran2020noisier2noise, batson2019noise2self, yaman2020self} that systematically avoid overfitting.

As mentioned above, for low-field MRI, it is often challenging or sometimes even impossible to obtain high-quality training data and/or paired image data. Therefore, self-supervised training approaches can be of high interest for this application. \\

The \textit{Noise2Noise} framework \cite{lehtinen2018noise2noise} relies on different realizations of noisy samples where the noise needs to be statistically the same between different samples. Noise2Noise is an attractive approach when different realizations of the same object can be easily acquired. For example, the approach was adopted in \cite{Lian2022ismrm}, where a cascaded model of sparsity transforms is used to reduce noise in images obtained on a $0.5\,\mathrm{T}$ scanner. \\

Noise2Noise requires different measurements of the same object, which can be challenging for MRI due to physiological changes occurring between scans. The \textit{Noisier2Noise} approach \cite{moran2020noisier2noise} overcomes this limitation by adding additional noise to the considered sample and using the original noisy sample as target data for training. At inference time, a scalar correction needs to be applied, which compensates for the learned ability of the network to reduce noise that is higher than the target noise of interest. A later discussed example of the application of Noisier2Noise to low-field MRI was presented in \cite{schlemper2021unsupervised}.\\

Finally, in \textit{Noise2Self} \cite{batson2019noise2self}, network training for the task of image-denoising is carried out by self-supervision employing the same target data that is used as network input. However, to avoid overfitting, the image pixels are split into two disjoint index sets, i.e.\ for $\boldsymbol{\rho}^{\eta}:=\boldsymbol{\rho}_{\mathrm{true}} + \boldsymbol{\eta} \in \mathbb{R}^N$, one chooses a partition of the index set $I=\{1,\ldots, N\}$, i.e.\ $I:=I_1 \cup I_2$ with $I_1\cap I_2 = \emptyset$ and defines the image $\boldsymbol{\rho}_{I_1}$ by  $[\boldsymbol{\rho}^{\eta}_{I_1}]_j = [\boldsymbol{\rho}^{\eta}]_j$ for $j \in I_1$ and $[\boldsymbol{\rho}^{\eta}_{I_1}]_j = 0$ for $j \notin I_1$. The input of the network is then restricted to pixels indexed by $I_1$ and the output of the network is compared to the input image restricted to pixels indexed by $I_2$.\\

\begin{figure}
    \centering
    \includegraphics[width=\linewidth]{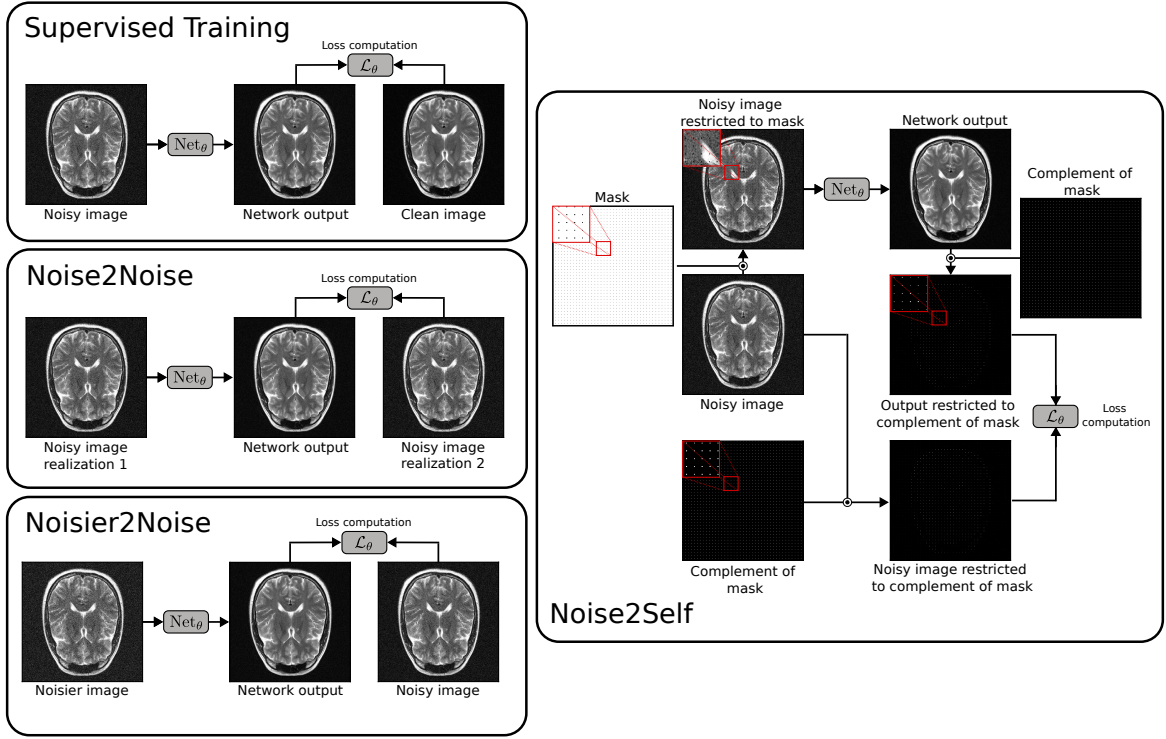}
    \caption{Different training techniques for a simple image-denoising example. Supervised training requires access to input-target image pairs and the network $\mathrm{Net}_{\theta}$ is trained to estimate the clean image from the noisy input. In the self-supervised Noise2Noise framework \cite{lehtinen2018noise2noise}, two noisy samples with identically distributed noise are required. For Noisier2Noise \cite{moran2020noisier2noise}, the input sample is generated by further corrupting the given noisy sample with additional noise. For Noise2Self \cite{batson2019noise2self}, the input image is restricted to a pre-defined mask and the estimated output is compared against the same input image projected on the complement of the employed mask. The principle of the framework self-supervision by data undersampling (SSDU) \cite{yaman2020self} (not shown in this Figure) that can be used for MR image reconstruction is similar to Noise2Self with the difference that the mask is applied in the Fourier domain of the considered image.}
    \label{fig:training_techniques}
\end{figure}

Figure \ref{fig:training_techniques} summarizes the just described schemes for a simple image-denoising example.

In \cite{yaman2020self}, the authors propose a self-supervision method that is closely linked to the above-discussed Noise2Self and Noisier2Noise, termed \textit{self-supervision by data undersampling} (SSDU). More precisely, the method aims at training a network to reconstruct images from accelerated MR data acquisitions where the $k$-space data in Eq.\,\eqref{eq:forward_problem} is undersampled. Thereby, similar to  Noise2Self, the (already undersampled) $k$-space data $\sd_J$ is split into two disjoint sets $J:=J_1\cup J_2$ with $J_1\cap J_2 = \emptyset$ and, by for example again choosing the squared $\ell_2$-norm as discrepancy measure, one can train the network by minimizing a loss function of the form
\begin{equation}\label{eq:loss_function_ssdu}
    \mathcal{L}_{\mathrm{SSDU}}(\theta) := \frac{1}{N_{\mathrm{train}}}\sum_{i=1}^{N_{\mathrm{train}}} \| \Ed_{J_2} f_{\theta}(\sd_{J_1}^{(i)}) - \sd_{J_2}^{(i)}\|_2^2.
\end{equation}
Further, in \cite{millard2023theoretical}, the connection between SSDU and Noisier2Noise is theoretically investigated.  We further also note that differences exist in the specific way of partitioning the data. For example, in \cite{yaman2020self} and \cite{millard2023theoretical}, the employed loss function resembles the Noisier2Noise framework, i.e.\ during training, the network learns to reduce artefacts from a higher undersampling than the one that is considered at test time. In contrast, in \cite{darestani2022test}, the distribution of the undersampling masks during training matches the one of the mask used for inference.

\begin{figure}
    \centering
    \includegraphics[width=\linewidth]{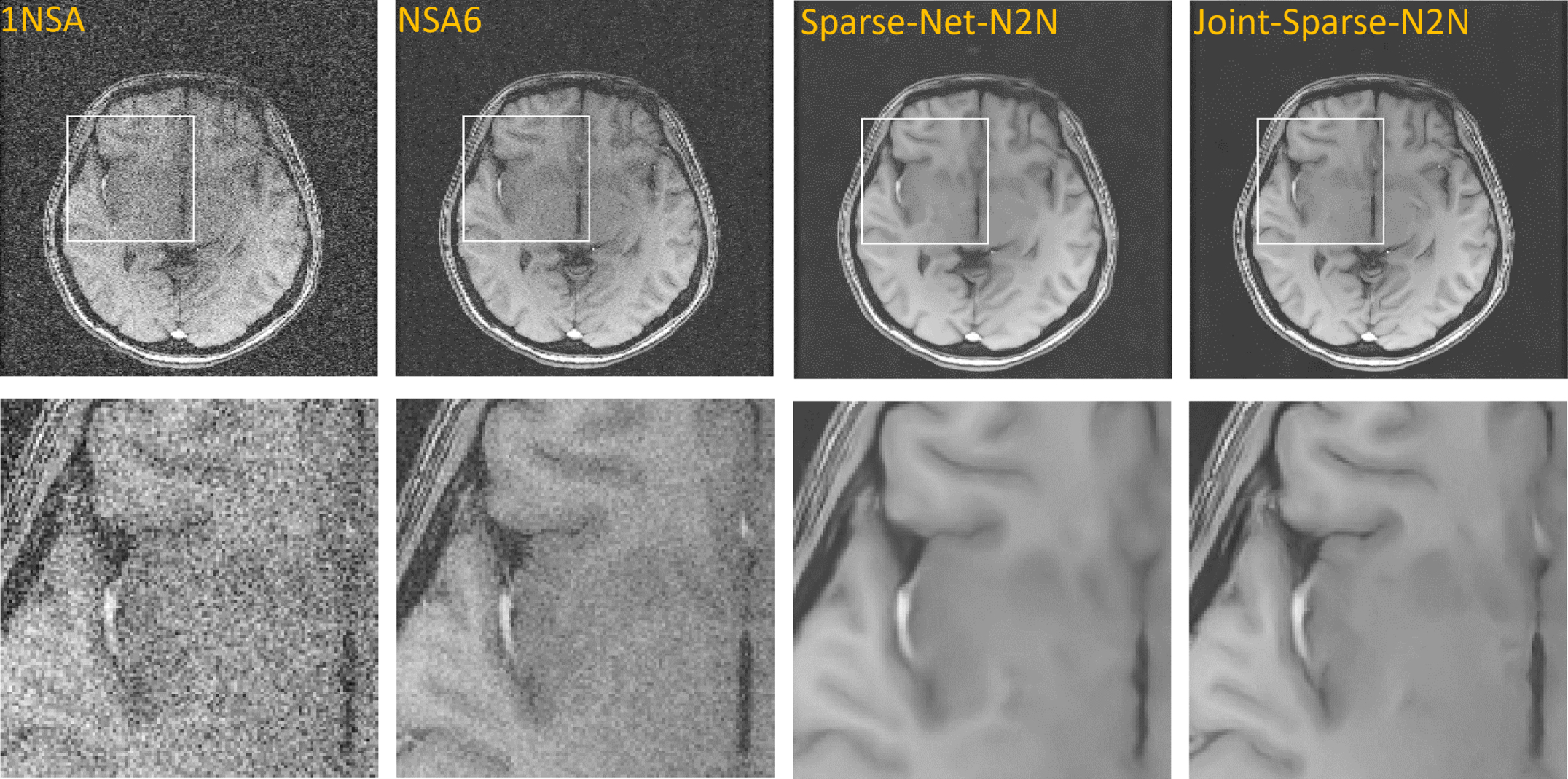}
    \caption{An example of images reconstructed with two deep neural networks for image denoising (Sparse Net and Joint Sparse-Net) that were trained using the Noise2Noise (N2N) framework on low-field MR image data acquired at $0.5\,\mathrm{T}$. The obtained images show a strong noise reduction compared to the images reconstructed from one and six averaged acquisitions (1NSA and NSA6), respectively. The image was adapted from \cite{Lian2022ismrm} and reproduced with the permission of the ISMRM. Image courtesy of Hua Guo.}
    \label{fig:denoising_example_lian2022ismrm}
\end{figure}

Figure \ref{fig:denoising_example_lian2022ismrm} shows an example of results that can be obtained by training an image-denoising network by self-supervised training on solely noisy image data using the Noise2Noise approach with the networks proposed in \cite{Lian2022ismrm}.

In the previously described methods, although no target images are assumed to be given, the workflow is still to train on a relatively large corpus of data and \textit{then} to apply the obtained model to new unseen samples.

Here, we also mention the possibility of applying so-called \textit{zero-shot methods}, i.e.\ where no dataset is used for training, but instead, the parameters of the model are adapted to the considered sample one aims to reconstruct, e.g.\ the well-known deep image prior (DIP) \cite{ulyanov2018deep}, \cite{yoo2021time}, the deep decoder \cite{arora2020untrained}, \cite{heckel2018deep}, \cite{darestani2021accelerated} as well as methods that employ self-supervised loss functions, see \cite{yaman2021zero}, \cite{lequyer2022fast}, \cite{mansour2023zero}. Also, see \cite{akccakaya2022unsupervised} for an extensive review of unsupervised deep learning-based methods for image reconstruction and image enhancement. 

Thereby, the idea is to reparameterize the sought image by a neural network, i.e.\ to set $\boldsymbol{\rho}:= f_{\theta}(\ZZ)$ for some (typically randomly chosen) input vector $\ZZ$. Then, instead of minimizing a variational problem of the form Eq.\,\eqref{eq:variational_problem}, one directly approaches the problem
\begin{equation}\label{eq:dip_problem_formulation}
    \min_{\theta} \frac{1}{2}\| \Ed f_{\theta}(\ZZ) - \sd\|_2^2
\end{equation}
for each new considered measurement $\sd$ with one of the previously described gradient descent-based methods, i.e.\ the loss-function corresponds to the data-discrepancy term and the employed dataset consists of only one sample.
Typically, however, as observed by the authors in \cite{ulyanov2018deep},  overfitting occurs eventually while minimizing Eq.\,\eqref{eq:dip_problem_formulation}, meaning that at some point, the network $f_{\theta}$ is able to approximate the noise present in the measurement $\sd$, thus not providing any implicit regularization effect. Thus, it is typically necessary to employ some regularization technique, e.g.\ early stopping as in the original paper \cite{ulyanov2018deep}, underparameterizing the network $f_{\theta}$ \cite{heckel_deep_2018} or combining DIP with regularization methods, e.g.\ with low-rank methods \cite{Hamilton2023magma} or with total variation-minimization (TV) approaches \cite{liu2019image}.

\begin{figure}
    \centering
    \includegraphics[width=0.98\linewidth]{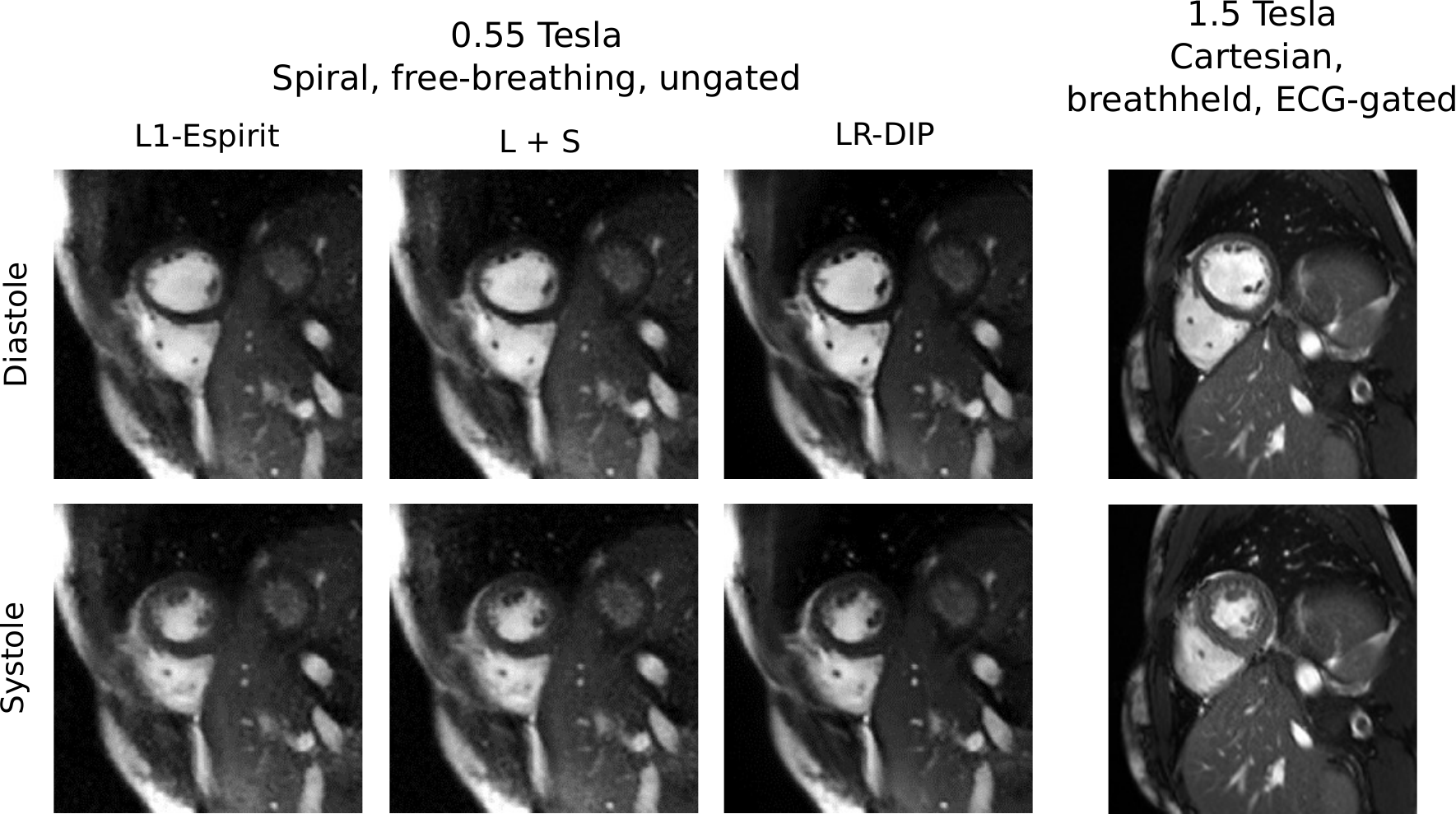}
    \caption{An example of cine images acquired at $0.55\,\mathrm{T}$ using spiral acquisitions during free breathing and reconstructed with $\ell_1$-ESPIRIT \cite{uecker2014espirit}, a low-rank + sparsity-based method (L+S) \cite{otazo2015low}, the proposed low-rank deep image prior (LR-DIP)\cite{Hamilton2023magma} and a reference image of the same subject obtained at $1.5\,\mathrm{T}$ acquired during a breathhold with an ECG-gated acquisition. The neural networks-based approach LR-DIP improves image quality over the two compressed sensing-based approaches and yields visually improved noise and artefacts suppression. The image was adapted from \cite{Hamilton2023magma}. Image courtesy of Jesse Hamilton.}
    \label{fig:lrdip-results}
\end{figure}

In  \cite{Hamilton2023magma}, for example, the DIP method combined with a low-rank (LR) approach was used to reconstruct cardiac cine data obtained at $0.55\,\mathrm{T}$. Figure \ref{fig:lrdip-results} shows an example of different reconstructions. The cine images obtained with the proposed LR-DIP show significantly higher image quality compared to a compressed sensing-based reconstruction ($\ell_1$-ESPIRIT) \cite{uecker2014espirit} and a low-rank + sparsity-based (L$+$S) \cite{otazo2015low}.

\subsubsection{Learned Reconstruction Methods}

After having introduced the basic notions of neural networks and their training, we are ready to discuss different strategies that can be employed for image reconstruction with deep learning-based methods in low-field MRI.

One of the perhaps most intuitive approaches is to construct a neural network $f_{\theta}$ that corresponds to a \textit{fully learned} mapping from the observed data to the unknown target image, i.e.\ to have a learned reconstruction mapping $\Ed_{\theta}^\sharp$ with $\Ed_{\theta}^\sharp(\sd)\approx \boldsymbol{\rho}_{\mathrm{true}}$. Thereby, the method is fully learned in the sense that one explicitly abstains from incorporating knowledge about the considered problem (e.g. the acquisition model $\Ed$ in \refeq{eq:forward_problem}) but instead aims at learning the entire mapping from the available training data. This approach is particularly suitable when one lacks a deep understanding of the forward model, which however does usually not apply to MR applications. For MRI, a fully learned inversion method corresponds to a function that directly maps the measured $k$-space data to an estimate of the ground-truth image without making use of the forward/inverse Fourier transform, for example. 

\begin{figure}
    \centering
    \includegraphics[width=0.95\linewidth]{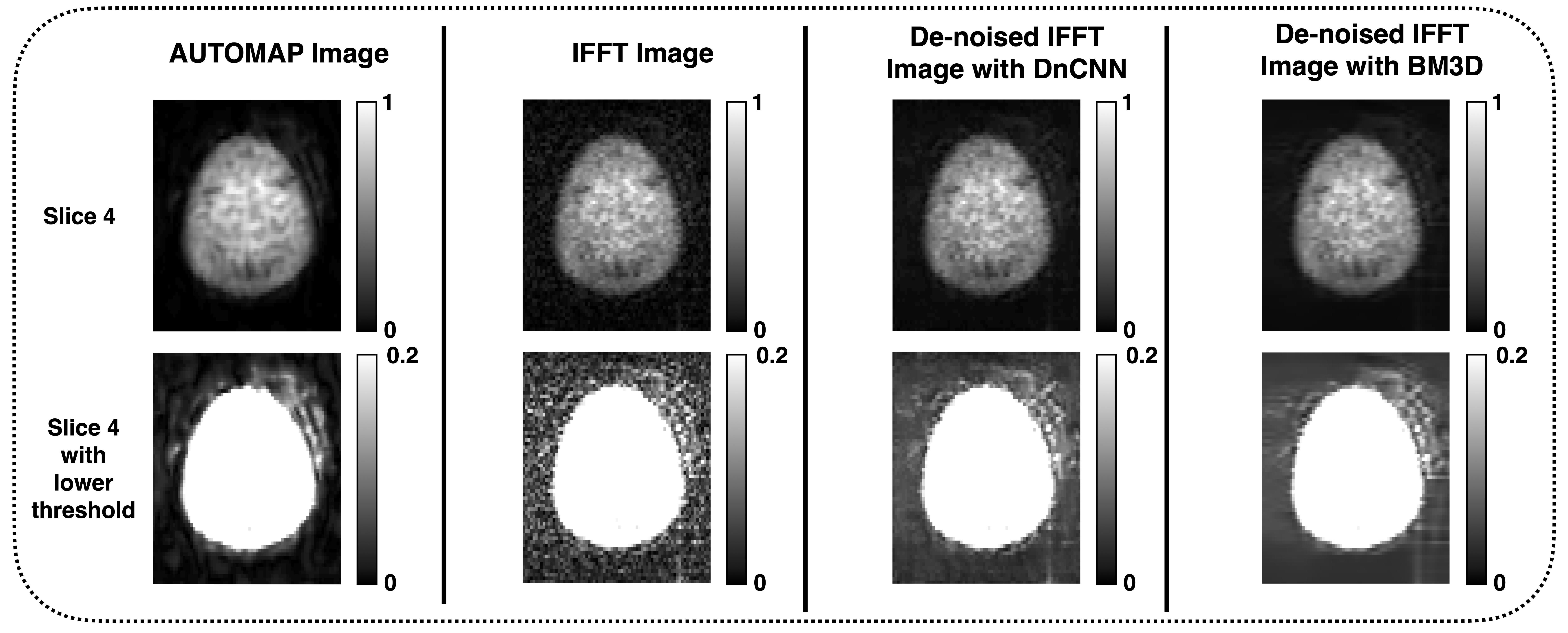}
    \caption{An example of ultra-low-field ($6.5\, \mathrm{mT}$) MR images of the brain reconstructed with different methods. From left to right: AUTOMAP \cite{zhu2018image, koonjoo2021boosting}, simple inverse FFT, inverse FFT image denoised with DnCNN \cite{zhang2017beyond}, inverse FFT image denoised with block-matching 3D \cite{dabov2006image, makinen2020collaborative}. Figure adapted from \cite{koonjoo2021boosting}. Image courtesy of Neha Koonjoo.}
    \label{fig:automap}
\end{figure}

In \cite{zhu2018image}, the authors proposed AUTOMAP, which allows for the inversion of arbitrary forward models using the same network architecture, provided that there exist input-target pairs to train the model on. Specifically for ultra-low-field MRI, AUTOMAP was applied in the work \cite{koonjoo2021boosting}, where the authors demonstrated the capability of the AUTOMAP architecture to invert 3D $k$-space data of the human brain measured at $6.5\, \mathrm{mT}$ using a bSSFP sequence, see Figure \ref{fig:automap}. The authors compared their method to two other well-known denoising methods that were applied after a simple inverse Fourier transform of the $k$-space data, namely block matching 3D (BM3D) \cite{dabov2006image} with collaborative filtering \cite{makinen2020collaborative} and the deep learning-based denoiser DnCNN \cite{zhang2017beyond} and found that AUTOMAP surpassed the methods of comparison both in terms of visual image quality by preserving more image details.

Fully learned inversion methods can be applied to super-resolution problems as well. Thereby,  to invert the low-resolution operator $\Rd$ in Eq.\,\eqref{eq:forward_operator_super_resolution}, the most common strategy is to first obtain an initial estimate of $\Rd \boldsymbol{\rho}_{\mathrm{true}}$, i.e.\ to invert the model $\Ed$ and then to train a network to estimate the underlying $\boldsymbol{\rho}_{\mathrm{true}}$. 

\begin{figure}
    \centering
    \includegraphics[width=0.95\linewidth]{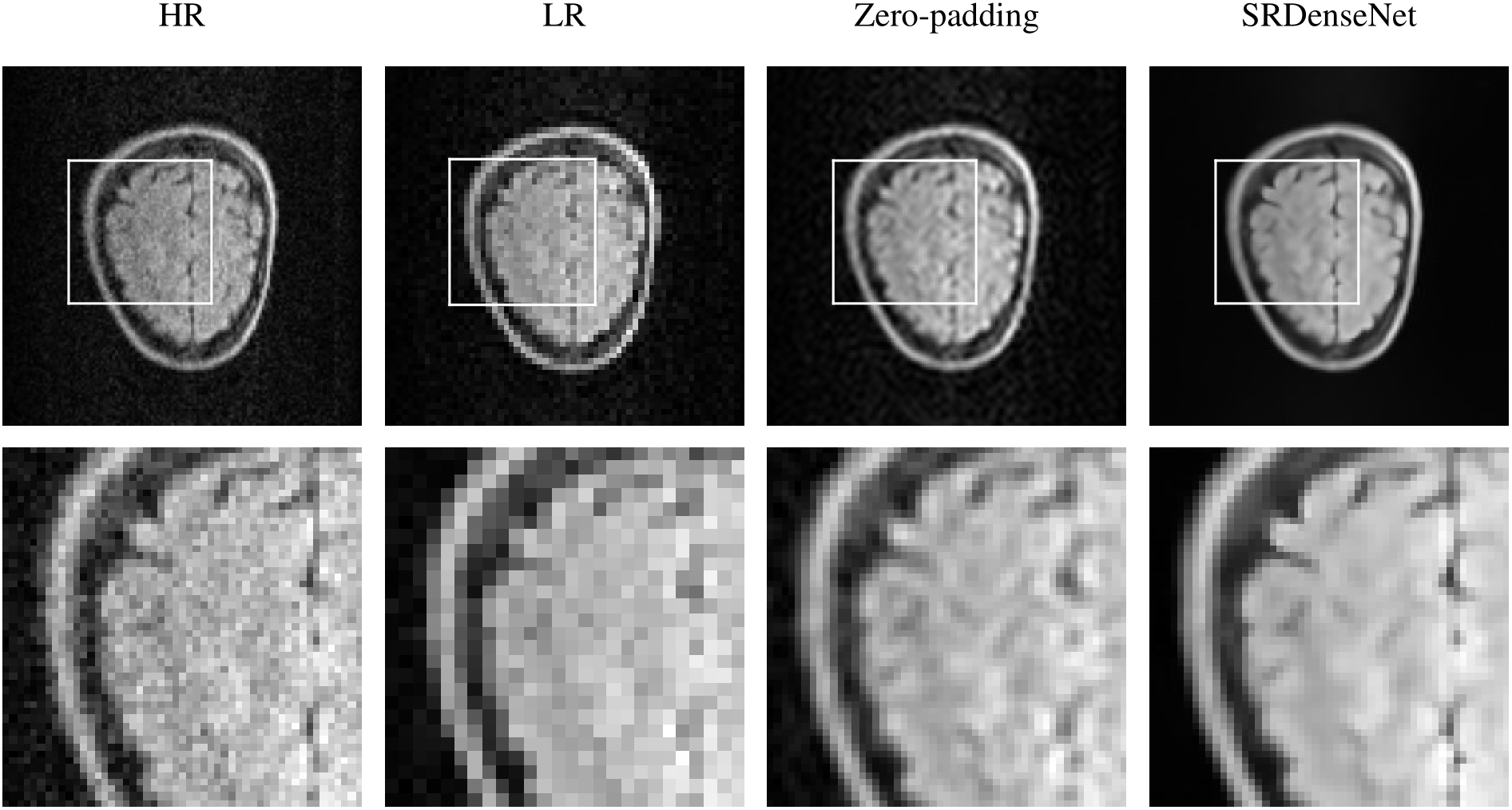}
    \caption{Example results obtained via a super-resolution network at ultra-low-field ($50\, \mathrm{mT}$). The image HR denotes a high-resolution slice of a brain image acquired an ultra-low-field MRI system, which serves as the reference image. The second image shows the low-resolution image simulated from the HR image using a sub-sampling forward model, see \cite{de2022deep} for details.  The third image is obtained by inverting the zero-padded $k$-space data with a simple IFFT and the fourth is the result generated by the proposed super-resolution network (SRDenseNet).  Figure adapted from \cite{de2022deep}. Image courtesy of Andrew Webb.}
    \label{fig:super_resolution}
\end{figure}

In \cite{de2022deep}, for example, the authors trained a CNN with dense skip connections termed Super-Resolution Dense Network (SRDenseNet), see \cite{tong2017image}, to map low-resolution images to their high-resolution counterpart by directly inverting the low-resolution operator, see Fig. \ref{fig:super_resolution}. Thereby, the target images were taken from the fastMRI brain dataset and the low-resolution images were retrospectively simulated by applying the four-fold low-resolution operator and further corrupting them with noise. The authors reported a noticeable improvement compared to the images obtained by zero-padding $k$-space data.

Another similar approach for portable ultra-low-field brain MRI at $64\, \mathrm{mT}$ \cite{iglesias2022quantitative} investigated the possibility of directly inverting the sub-sampling operator using a U-Net following the approach in  \cite{iglesias2021joint}, which also yields image segmentations of 39 regions of interest of the brain. Thereby, the obtained results on their brain MR image data were also evaluated in terms of the ability of different segmentation tools to generate accurate region-of-interest volumes from the obtained images. The authors reported good agreements between the segmentations obtained from the reference high-field strength and the ones obtained from their learned upsampling method, which demonstrates the applicability of these methods with respect to subsequent downstream tasks, e.g.\ segmentation.

In super-resolution applications, the networks are usually trained on retrospectively simulated data, i.e.\ high-resolution images are corrupted by down-sampling to generate the input-target image pairs to be used for training. In \cite{laguna2022super}, the authors further explored a method that employs a domain-adaptation module based on unpaired image translation from brain MR images acquired at $64\, \mathrm{mT}$ to high-field images obtained from the Connectome project \cite{van2013wu}. Their method comprises a GAN \cite{zhu2017unpaired} and a patch-based contrastive learning approach \cite{park2020contrastive}. Since these two modules can be trained in an unsupervised fashion, the approach presented in \cite{laguna2022super} eliminates the need to have access to paired image data. Similar to the previously discussed approaches, the method aims at inverting the low-resolution operator $\Rd$ in Eq.\,\eqref{eq:forward_operator_super_resolution}.

Fully learned inversion methods, however, have, first of all, the inherent conceptual limitation of ignoring all available and valuable information that the (at least partially) known physical model provides. Second, they might lack data consistency, i.e.\ it is not clear how well the estimated quantities match the observed data when the forward model is applied. Third, since they typically consist of at least a few fully connected layers (i.e.\ some of the matrices $\mathbf{W}$ in Eq.\,\eqref{eq:neural_network} are dense matrices) to allow for high expressiveness to be able to learn to invert the forward model, their application is limited to input data that has the same dimension as the data the network was trained with. This, in contrast, is not the case for networks consisting of solely convolutional layers and thus represents a limitation from a practical point of view as either a new network has to be trained for each input dimension or data must be pre-processed accordingly.  Additionally, although the authors in \cite{zhu2018image} and \cite{koonjoo2021boosting} report that AUTOMAP is inherently robust, other works have shown that AUTOMAP was the least robust compared to other methods with respect to adversarial attacks \cite{antun2020instabilities}, especially compared to deep learning methods that employ the data acquisition model as part of their architecture. 

\textit{Post-processing reconstruction}  or \textit{image enhancement} methods, as opposed to fully learned methods, do not entirely ignore the structure of the given problem at hand at inference time. Instead, one typically obtains an initial estimate of the unknown ground-truth image from the noisy and/or undersampled $k$-space data, by applying some reconstruction operator, e.g.\ the adjoint $\Eu^\herm$ or the pseudo-inverse $\Ed^\dagger$, and subsequently trains a network to reduce the resulting noise and artefacts. As discussed in Chapter 7, this type of approach has been extensively investigated for MRI in a large number of works for different organs (e.g.\ knee \cite{jin2017deep, ding2019deep}, brain \cite{wang2016accelerating, hyun2018deep, lee2018deep, kawamura2021accelerated, li2020mri}, cardiac \cite{kofler2019spatio, Hauptmann2019, sandino2017deep}) and has shown promising results. Additionally, we note that there exists the alternative, where, instead of learning the denoising/artefacts reduction mapping in the image space, one can learn a suitable transformation for filling the $k$-space and subsequently applying a simple inverse Fourier transform \cite{cheng2018deepspirit, akccakaya2019scan, han2019k}.

Specifically for low-field MRI, the authors in \cite{ayde2022deep} used a 2D U-Net with residual connections for image denoising and artefacts reduction from undersampled low-field MRI scans acquired at  $0.1\, \mathrm{T}$, where, by the use of data-augmentation techniques, they were able to successfully train the network even on a relatively small dataset only comprising 10 human wrist images.

Further, the authors in \cite{le2021deep} proposed a method for simulating scanner-specific images from the publicly available dataset of the Cancer Imaging Archive \cite{clark2013cancer} as well as a post-processing reconstruction method for reducing different types of artefacts related to B0-inhomogeneities, nonlinear gradients, and undersampling. The post-processing reconstruction method, which is based on a stacked U-Net architecture, was then applied to images obtained from an ultra-low-field Promaxo MR scanner with a low and non-uniform B0 field strength of $60-67\,\mathrm{mT}$.

In another work \cite{kojima2022denoising}, the authors investigated the applicability of the Noise2Void \cite{krull2019noise2void} framework to train a denoising U-Net on patches of T1- and T2-weighted images of a kiwi that was scanned as a phantom on a $0.35\,\mathrm{T}$ MRI system.

Another approach \cite{lucas2023multi} employs a generative adversarial network (GAN) that was learned on paired brain MR image data acquired at $64\,\mathrm{mT}$ and $3\,\mathrm{T}$, respectively. The authors evaluated the obtained synthesized images with respect to image quality, brain morphometry, and white matter lesions and found that quantitative measures such as thalamic, lateral ventricle, and total cortical volumes in the synthesized images did not significantly differ from the images obtained at $3\,\mathrm{T}$.

In \cite{schlemper2021unsupervised}, the authors explored the previously discussed Noisier2Noise framework to learn to denoise an image in an self-supervised fashion. The work in \cite{schlemper2022two} followed a two-step approach by first pre-training a DnCNN denoiser in a supervised fashion on a large corpus of data given by the Human Connectome Project (HCP) \cite{van2013wu}. Then, in a second step, the pre-trained is applied to the noisy DWI images obtained from a $64\, \mathrm{mT}$ scanner to obtain pseudo-labels to be added to the initial set  for a new training that is performed from scratch using the extended dataset. 

One main challenge of employing deep neural networks that are trained in a supervised manner is that they must exhibit good generalization properties. Typically, since they contain a large number of trainable parameters, the dataset used for training must be rather large. However, for some applications, e.g.\ low-field MRI, obtaining high-quality target images is simply not feasible. Thus, one possibility is to employ one of the previously described self-supervised training techniques. Another option is to rely on the availability of high-quality image data obtained from MR scanners at higher magnetic fields and to retrospectively simulate the required data pairs. However, depending on one's capability to accurately simulate the image degradation process the network's performance at test time might suffer more or less from inherent distribution shifts between the training and the target test data.

\begin{figure}
    \centering
    \includegraphics[width=0.95\linewidth]{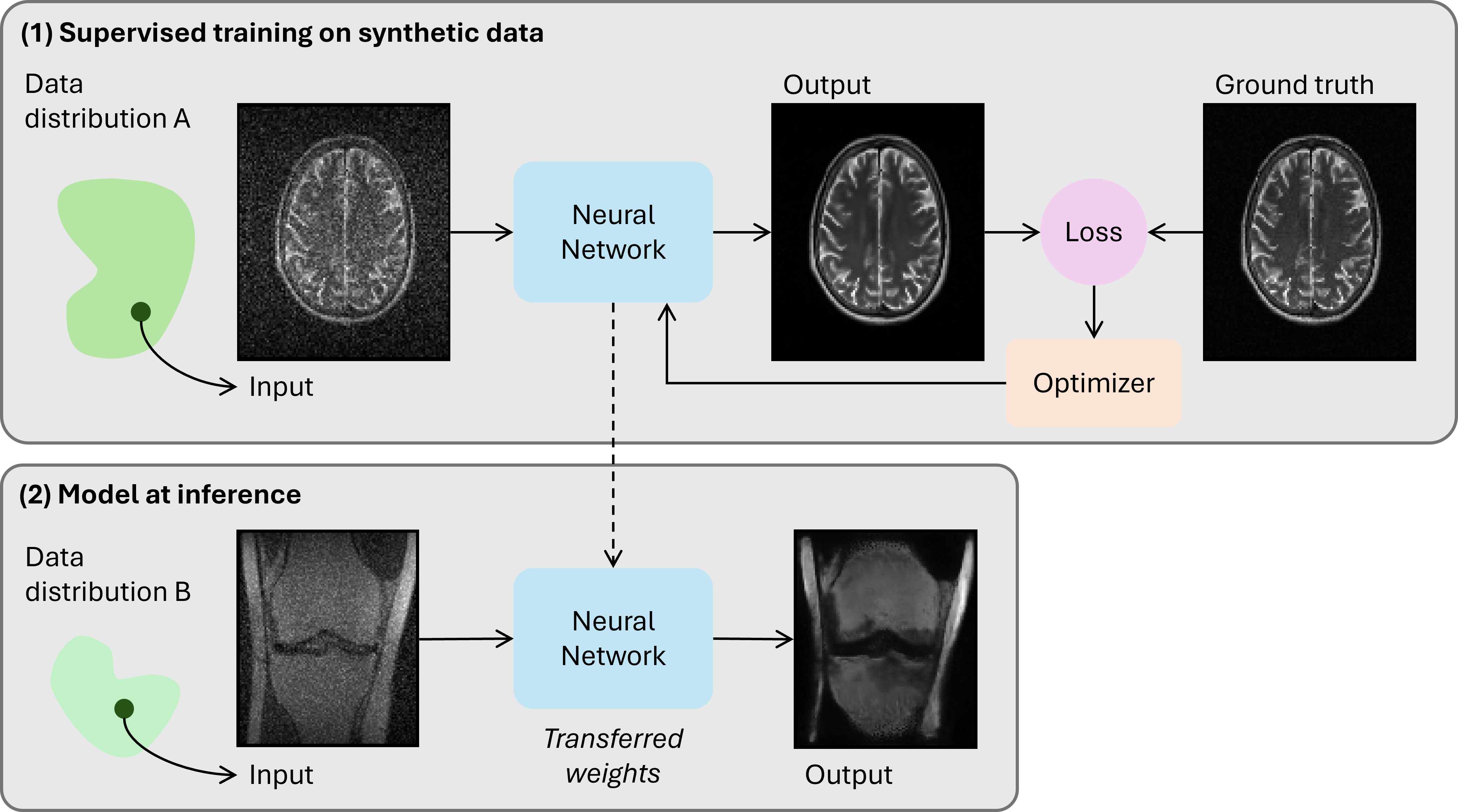}
    \caption{Training data for low-field MRI denoising problems is usually rare. Therefore, many approaches use retrospectively simulated low-field data to train models and transfer the results to real data. If the two data distributions do not match (i.e.\ there is a noticeable distribution shift) the output of the network can lead to severe artefacts. In this example, applying a denoising network that was trained solely on $T_2$-weighted brain images and then applying the obtained model to $T_1$-weighted knee images leads to severe signal voids within the bone marrow.}
    \label{fig:denoising_experiments_test_time_training}
\end{figure}

Another case where such a performance decrease attributable to a distribution shift is observable is when training and test data differ from each other in terms of anatomy and image contrast. Figure \ref{fig:denoising_experiments_test_time_training} demonstrates this on an image-denoising example for low-field data. A denoising network is trained on T2-weighted brain images and the network is then directly applied to T1-weighted knee images. Here, training and test data strongly differ due to their different contrast as well as the different anatomy, leading to severe signal voids in the denoised knee images. We point out that this example is extreme in the sense that on average, such drastic examples are fortunately much rarer to be found, but it highlights the importance of exposing the network to a large variety of different images. So-called test-time-training \cite{darestani2022test} can help close the performance gap that is typically observed when applying neural networks to highly different test data. A recent study shows that, in general, to achieve a small gap between the performance on different datasets in the first place, it seems to be beneficial in terms of robustness to employ rich and diverse training data, even when the target data is only a sub-class of the data used for training \cite{Lin2024Robustness}.

Post-processing reconstruction methods, however, have one major limitation. Because from a theoretical point of view, the exact functioning mechanism of neural networks is not well understood, this type of approach rightfully raises the question of how reliable they can expected to be. These concerns are further underpinned by some works that have reported hallucinations of deep neural networks for the task of image reconstruction \cite{Muckley2021}. Further details regarding the robustness of deep learning reconstruction and hallucinations can be found in Chapter 11. 

The last mentioned limitation of image enhancement methods can partially be addressed by employing reconstruction methods that employ neural networks only as one of several distinct steps, often referred to as \textit{model-unaware} or \textit{decoupled methods}. For example, when aiming to regularize inverse problems, so-called Plug and Play (PnP) methods \cite{venkatakrishnan2013plug} or regularization by denoising (RED) methods \cite{romano2017little} are image reconstruction frameworks that alternate between steps motivated from model-based reconstruction and the application of general Gaussian denoisers, which can, in particular, be given by state-of-the-art deep neural networks such as the well-known DnCNN \cite{zhang2017beyond}. 

These methods are attractive for one main reason. Training the Gaussian network denoiser is completely decoupled from the considered image reconstruction process, which makes it computationally relatively cheap to accomplish. Further, image denoising is, due to its inherent structural simplicity, without doubt, one of the most well-studied inverse problems and has thus attracted large attention of the signal processing community, yielding a large amount of available "off-the-shelf" Gaussian denoisers that can be directly integrated into the PnP or RED framework. Compared to purely post-processing-based reconstruction methods, they in addition increase the data consistency of the obtained solution since the non-learned algorithm components typically involve the use of the measured $k$-space data as well as the adjoint/forward operator. Additionally, this type of method was also reported to be applicable with networks that reduce artefacts instead of Gaussian denoisers, see e.g.\ \cite{liu2020rare}.

Interestingly, to the best of our knowledge, these types of methods, i.e.\ PnP or RED, have not been applied to low-field or ultra-low-field MRI yet, although they naturally fit well with the given low SNR regime. 

Further, we note that model-unaware or decoupled methods can not only be employed to learn artefacts reduction- or denoising mappings but in general to estimate any relevant quantity to be used within the image reconstruction pipeline. For ultra-low-field MRI, field inhomogeneities leading to image distortions are often challenging. Therefore, approaches that estimate and correct for these inhomogeneities are of specific interest.

For instance, the work in \cite{daval2023deep}, although not specifically designed for low-field MRI, but rather for non-Cartesian susceptibility-weighted imaging (SWI), which also suffers from field-inhomogeneities induced by the patients during long readouts, proposed a reconstruction method that compensates for the field-inhomogeneity correction utilizing a neural network. The studied neural network architectures consist of reconstruction methods based on the non-Cartesian primal-dual network architecture \cite{ramzi2022nc}. For the training, they employ target images that were obtained by a self-corrected method \cite{daval2022iterative}. The network is trained in a so-called greedy fashion, i.e.\ the learnable components of the networks are trained using the current estimates of the reconstruction algorithm as input images. 

Other works \cite{zeng2019off}, albeit not for low-field MRI applications but for pediatric scanning, proposed a network termed Off-Resonance Correction Network (Off-ResNet) that can be used to estimate the field-inhomogeneity-corrected image map from an input image. Similarly, the authors in \cite{haskell2020deep} propose to employ the same network architecture but to train it to estimate the field-inhomogeneity map itself such that a subsequent correction with a B0-informed method \cite{fessler2005toeplitz} can be applied. 

A  third important category of state-of-the-art reconstruction algorithms using deep learning are so-called physics-informed methods based on algorithm unrolling \cite{monga2021algorithm}, see Chapter 7 and the review \cite{hammernik2023physics} for an in-depth discussion. In contrast to the methods described above, the learned networks themselves already constitute reconstruction algorithms that contain both learnable parameters as well as model-based blocks describing the physics of the considered problem. As such, the learned components are not only related to the data they were trained with but also to the reconstruction pipeline that the network defines. Thus, these networks are sometimes also referred to as physics-informed neural networks.

In \cite{schlemper2020deep}, the authors presented an end-to-end trainable reconstruction network that addresses undersampling and hardware artefacts for an ultra-low-field Hyperfine MRI scanner at $64\,\mathrm{mT}$. Their network consists of a 2D non-uniform variational network \cite{schlemper2019nonuniform} and shares similarities to the cascade presented in \cite{schlemper2017deep}. It was trained on images taken from the Human Connectome Project \cite{van2013wu}. Those images were retrospectively resampled to match the resolution of their Hyperfine T2-weighted fast spin echo images and field of view. Additionally, phase and coil-sensitivity weighting was introduced before retrospectively generating $k$-space data. They compared their method to a standard 2D U-Net post-processing reconstruction method as well as to a $k$-space-based U-Net \cite{han2019k}, and reported improvements. 

Similar studies have been carried out for $0.55\,\mathrm{T}$. In \cite{Schlicht2024ejr} a standard $k$-space-based denoising and super-resolution approach was compared to a variational network approach \cite{hammernik2018learning} for lumbar spine imaging at $0.55\,\mathrm{T}$. The variational network approach allowed for a scan time reduction of 40\% while ensuring high image quality compared to the standard method. The same variational network approach was also compared for knee imaging at $0.55\,\mathrm{T}$ \cite{Donners2024ir}. The variational network approach yielded comparable diagnostic confidence as the images obtained at $3\,\mathrm{T}$. Another study comparing knee imaging at $0.55\,\mathrm{T}$ and $1.5\,\mathrm{T}$ came to similar results \cite{schmidt2023jcm}.

Note, however, that although these methods incorporate the physical model in blocks of their network architecture, their considered model for example ignores B0-field inhomogeneities, which is an important challenge for ultra-low-field MRI.

In contrast, the work in \cite{schote2024joint} corresponds to an end-to-end trainable reconstruction method that estimates both the image as well as a field map describing the underlying B0-field inhomogeneities which are responsible for image distortions and which was applied to retrospectively simulated ultra-low-field knee MR data. Thereby, for the estimation of the field map, the authors use a neural network architecture that, instead of estimating the entire field map directly as in \cite{haskell2020deep}, estimates the coefficients of the spherical harmonic functions \cite{schote31physics, schote_reliable_2023} that are known to well-approximate the B0-field inhomogeneities \cite{wenzel2021b0}. 

\begin{figure}
    \centering
    \includegraphics[width=0.9\linewidth]{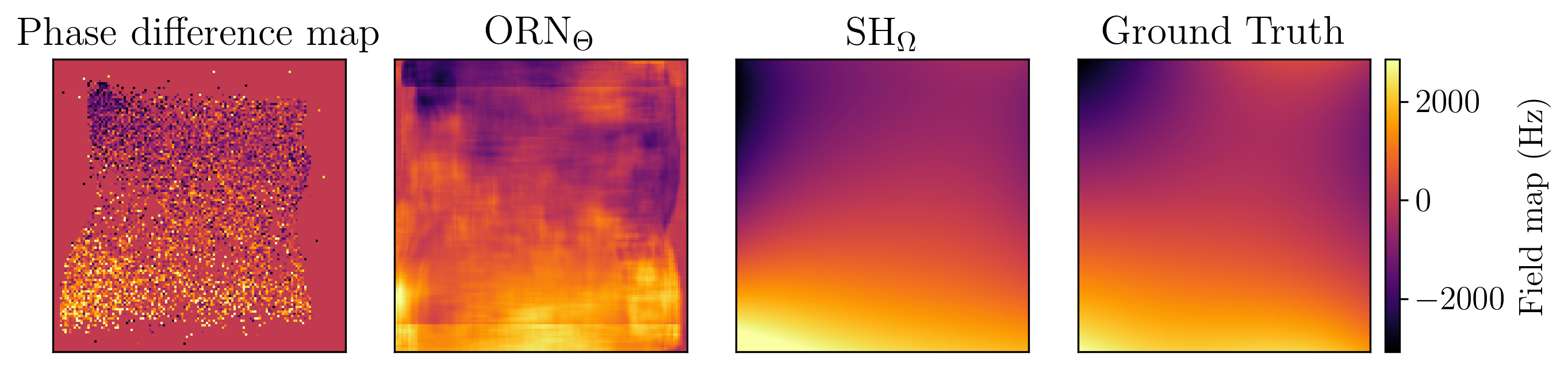}
    \caption{Results of different approaches for estimating the field map from low-field MRI data. The conventional phase difference map obtained from two acquisitions with a shift of the acquisition window results in a very noisy field map. The off-resonance net ($\mathrm{ORN}_{\Theta}$) \cite{zeng2019off, haskell2020deep} developed for higher field strengths combines different convolutional layers with residual connections but leads to patch-like artefacts in the field map. Using spherical-harmonic basis functions ($\mathrm{SH}_{\Omega}$) \cite{schote31physics, schote_reliable_2023} as a decoder and learning the spherical harmonic coefficients restricts the solution to a smooth representation. Image adapted from \cite{schote2024joint}.}
    \label{fig:field_map_results}
\end{figure}

Figure \ref{fig:field_map_results} shows a comparison of the field map estimated from a noisy and distorted input low-field image using the previously mentioned Off-Resonance Network (ORC-Net) and the Spherical Harmonic Network (SH-Net) proposed in \cite{schote2024joint}. Additionally, the field map that is estimated from the phase difference of two different simulated low-field MR acquisitions is shown. As we can see, the latter is highly non-smooth and contains a noticeable amount of noise, while the ones estimated by the CNNs are superior in terms of smoothness. Further, the one calculated by the SH-Net is smoother than the one obtained by the ORC-Net as the output of the network is constructed to be a linear combination of spherical harmonic functions up to a certain order.

In \cite{schote2024joint}, a second U-Net serves as a denoising network and its output is used to regularize the problem employing a Tikhonov regularization term $\mathcal{R}(\boldsymbol{\rho}):=\|\boldsymbol{\rho}- \boldsymbol{\rho}_{\mathrm{NN}}\|_2^2$, as used in many other works, e.g.\ \cite{aggarwal2018modl}, \cite{kofler2021end}. Thereby, the output $\boldsymbol{\rho}_{\mathrm{NN}}$ is obtained by applying the U-Net to the noisy image for which the distortion has been corrected by using the field map estimated by the previously discussed SH-Net. The entire physics-informed network in \cite{schote2024joint} was trained on retrospectively simulated data from the knee fastMRI dataset \cite{zbontar2018fastmri} and the authors reported improvements when comparing their method to other non-learned methods. This approach falls in the category of algorithm unrolling as the entire network resembles an iterative scheme for the joint estimation of the B0 field map and the image, but, in contrast to other works, e.g.\ \cite{usman2020joint}, uses learned regularization terms.  

\begin{figure}
    \centering
    \includegraphics[width=\linewidth]{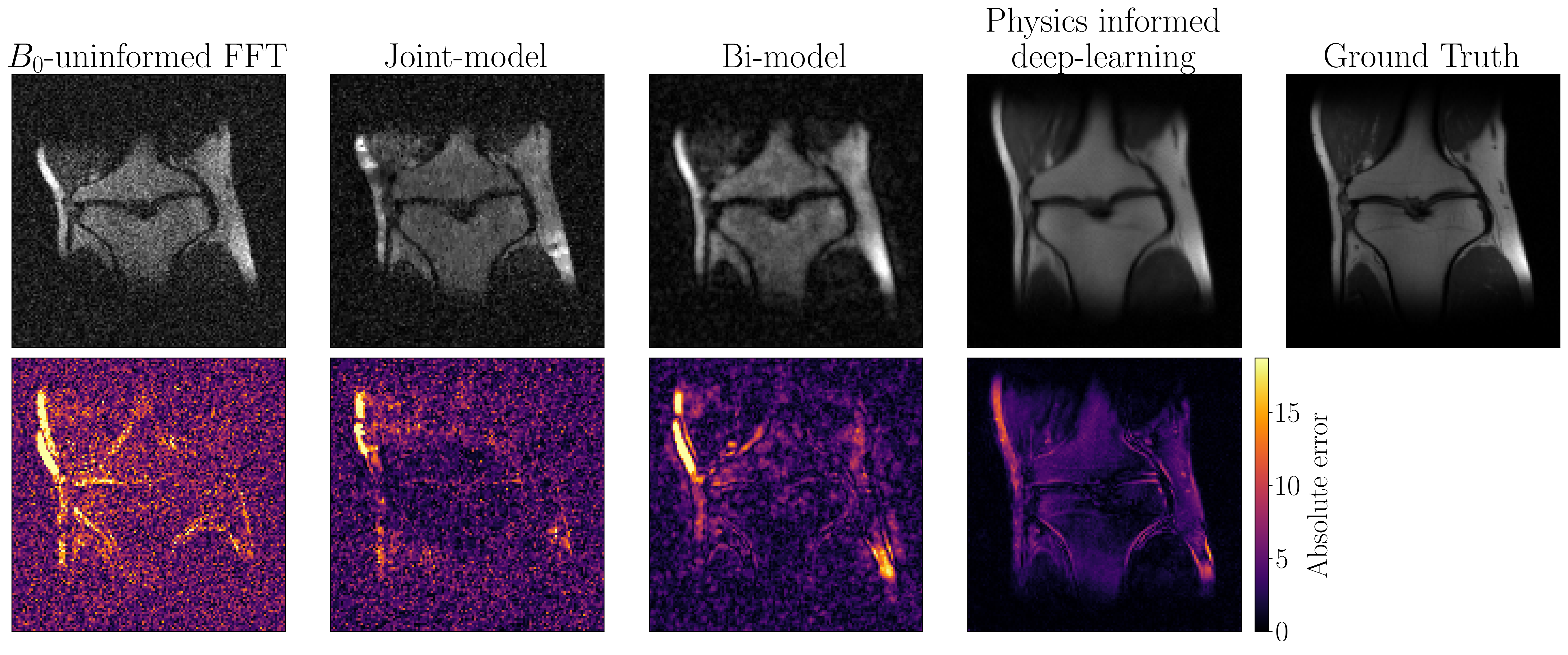}
    \caption{Comparison of different joint image and field-map reconstruction methods for simulated low-field data. The B0-uninformed IFFT reconstruction reveals the image distortion caused by B0-inhomogeneities if not corrected. The conventional approaches provide iterative reconstruction results by involving one joint model or two separate models. Due to the low SNR, the physics-informed deep learning approach results in better contrast and fewer image distortions.}
    \label{fig:joint_recon_results}
\end{figure}

Figure \ref{fig:joint_recon_results} shows a comparison of knee images from the fastMRI dataset \cite{zbontar2018fastmri} for which a retrospective low-field data acquisition was simulated, see \cite{schote2024joint} for details. The images were reconstructed from the noisy $k$-space data with the methods presented in \cite{usman2020joint} and \cite{koolstra2021image}, which both rely on non-learned image priors and the end-to-end trainable network \cite{schote2024joint}, which uses CNNs for the regularization of both the image and the field-map estimates. All three methods provide estimations of the field map and the image. 

\subsection{AI solutions for image processing in low-field MRI}
Last, we briefly discuss some recent works that explore opportunities given by deep learning methods for different image processing tasks, e.g.\ image segmentation or image classification.

An important quality control step in clinical practice is the identification of artefacts in the obtained images. In addition to detecting artefacts, it is also important to correctly classify them as this allows the user to adapt the MR acquisition to reduce or avoid the artefact in a repeated acquisition. The authors in \cite{jimeno2022mri} presented an approach that detects and classifies Cartesian undersampling artefacts and artefacts due to Gibbs ringing for brain imaging at $0.36\,\mathrm{T}$. Their method also yields heat maps that highlight the regions in the image affected by artefacts. 

Automatic segmentation of different organs is of high interest to speed up medical examinations and allow for automatic assessment of diagnostic parameters. The authors in \cite{payette2023miccai} utilized a U-Net trained on high-field and low-field fetal image data to segment organs of interest (e.g.\ lungs, liver, spleen, kidneys). The method was evaluated on $0.55\,\mathrm{T}$ fetal images and used to automatically assess $\mathrm{T}2^*$ values in these organs as a biomarker for blood oxygenation.

In \cite{Folkesson2007tmi}, a segmentation approach for knee imaging at $0.18\, \mathrm{T}$ was developed based on a $k$-nearest neighbors algorithm trained on manual segmentations. Segmentation of tibial and femoral cartilage was achieved in healthy subjects and patients suffering from osteoarthritis. 

In \cite{bhattacharjee2023feasibility}, the authors explore the applicability of deep learning-based image segmentation models trained on $3\,\mathrm{T}$ knee MR image data to $0.55\,\mathrm{T}$ MR images. The results suggested that bone segmentations were in general slightly better at $3\,\mathrm{T}$, and the quality of the cartilage segmentations was comparable between the $0.55\,\mathrm{T}$ and $3\,\mathrm{T}$.

The work in \cite{bhattacharya2020automated} investigated the applicability of a Cycle-GAN-based approach \cite{zhu2017unpaired} for the segmentation of lung cysts at $0.55\,\mathrm{T}$ MRI. The employed dataset consisted of 65 lymphangioleiomyomatosis patients for images that were acquired on a $0.55\,\mathrm{T}$ MRI scanner as well as on a CT. Thereby, one part of the network first generates synthetic CT images from the MR images for which then a second module based on residual U-Nets segments lung volume and cysts. The network was trained on 12089 CT slices and 1258 MRI slices using an adversarial loss as well as a cycle-consistency loss.  The authors reported good agreements between the segmentations obtained from the CT images and the synthetic CT images generated from the low-field MR images, which shows the potential of employing an ionizing radiation-free alternative for the quantification of lung cysts.

\section{Conclusions}
The revival of low-field and ultra-low-field MRI presents an exciting opportunity to address key limitations in the accessibility, cost, and safety of MRI technology. While low-field MRI still faces significant challenges, particularly in SNR, spatial resolution, and scan efficiency, ML/AI-driven solutions hold the potential to overcome these barriers. By leveraging deep learning techniques for image reconstruction, denoising, automatic segmentation, and post-processing, low-field MRI systems can produce higher-quality images that approach the diagnostic capability of high-field systems. Moreover, ML-based innovations in acquisition planning and scan time further enhance the practicality of low-field MRI in clinical environments.

As ML technologies continue to evolve, their integration with low-field MRI has the potential to transform the landscape of medical imaging, making MRI more accessible and cost-effective worldwide. However, future research is needed to further advance these ML-based methods and establish regulatory standards that ensure consistent image quality and safety. Through continued innovation, low-field and ultra-low-field MRI could become vital tools in global healthcare, expanding access to critical diagnostic imaging, particularly in regions where cost and accessibility are major concerns and/or enabling timely assessments in point-of-care settings.

\bibliographystyle{vancouver}
\bibliography{references}
\end{document}